\pdfoutput=1
\documentclass[12pt]{article}

\usepackage{cite}
\usepackage{amsmath}
\usepackage{pstricks}
\usepackage{bbold}
\usepackage{slashed}
\usepackage{amssymb}
\usepackage{graphicx}
\usepackage{caption}
\usepackage{subcaption}
\usepackage{hyperref}
\usepackage{verbatim}
\usepackage{multirow}

\def\mathswitch#1{\relax\ifmmode#1\else$#1$\fi}
\def\mathswitchr#1{\relax\ifmmode{\mathrm{#1}}\else$\mathrm{#1}$\fi}

\newcommand{\tev}{\,\, \mathrm{TeV}}

\DeclareMathOperator*{\maxr}{Max}

\begin{document}
\thispagestyle{empty}

\def\thefootnote{\fnsymbol{footnote}}

\vspace{1cm}

\begin{center}

{\Large\sc {\bf  Supersymmetry with a Heavy Lightest Supersymmetric Particle  }}
\\[3.5em]
{\large\sc
Taoli Cheng$^{1,3}$\footnote{ctl@itp.ac.cn}, Jinmian Li$^1$\footnote{phyljm@gmail.com}, Tianjun Li$^{1,2}$\footnote{tli@itp.ac.cn}
}

\vspace*{1cm}

{\sl $^1$
State Key Laboratory of Theoretical Physics, Institute
of Theoretical Physics, Chinese Academy of Sciences, Beijing 100190,
P. R. China
}
\\[1em]
{\sl $^2$
School of Physical Electronics, University of
Electronic Science and Technology of China, Chengdu 610054, P. R.
China
}
\\[1em]
{\sl $^3$
Max-Planck-Institut f\"ur Physik (Werner-Heisenberg-Institut), \\
   F\"ohringer Ring 6,  80805 M\"unchen, Germany\\
}

\end{center}

\vspace*{2.5cm}

\begin{abstract}

To escape the current LHC supersymmetry (SUSY) search constraints while preserve 
the naturalness condition, we propose the heavy Lightest Supersymmetric Particle 
(LSP) SUSY. According to the different dependence on the LSP mass, we systematically 
classify the discriminating variables into three categories. We find the strong 
dependence of all current SUSY searches on variables in the first category render 
the weak sensitivity for the heavy LSP SUSY. Especially, all the current LHC SUSY 
search constraints can be evaded if the LSP mass is around 600 GeV or higher. In 
the Minimal Supersymmetric Standard Model (MSSM), we find that the heavy LSP SUSY 
does not induce more fine-tuning than Higgs boson mass. Moreover, the muon anomalous 
magnetic moment can be satisfied within 3-$\sigma$ level.  We systematically study 
the viable parameter space for heavy LSP SUSY, and present four benchmark points 
which realize our proposal concretely. An improved collider search for those benchmark 
points, which mainly relies on the variable in the second category, is discussed 
in detail.

\end{abstract}
\setcounter{page}{0}
\setcounter{footnote}{0}

\newpage


\section{Introduction}
\label{intro}

The Standard Model (SM)-like Higgs boson with mass $m_h$ around 125 GeV was discovered
at the first run of the LHC in July 2012~\cite{Aad:2012tfa, Chatrchyan:2012ufa}.
Such a large Higgs boson mass requires the multi-TeV top squarks with small mixing or TeV-scale 
top squarks with large mixing in the Minimal Supersymmetric SM 
(MSSM)~\cite{Nilles:1983ge, Haber:1984rc, Martin:1997ns}. Moreover, the null results for
the LHC supersymmetry (SUSY) searches have given strong constraints on
the pre-LHC viable parameter space. For instance, the gluino mass $m_{\tilde g}$ should
be heavier than about 1.7 TeV if the first two-generation squark mass $m_{\tilde q}$ 
is around the gluino mass $m_{\tilde q} \sim m_{\tilde g}$, and heavier than about 1.3 TeV
for $m_{\tilde q} \gg m_{\tilde g}$~\cite{Chatrchyan:2013wxa, Aad:2014wea}.
Because the colored supersymmetric particles (sparticles) should be above TeV scale for
the ordinary SUSY scenarios~\cite{lhc}, the little hierarchy 
problem~\cite{BasteroGil:2000bw, Bazzocchi:2012de} and naturalness problem~\cite{Papucci:2011wy} 
in SUSY models are aggravated.

Many new models have been proposed to increase the Higgs boson mass. For example,
in the singlet or triplet extended MSSM~\cite{Balazs:2007pf, Ellwanger:2009dp, Espinosa:1991wt, FelixBeltran:2002tb, DiChiara:2008rg}, the SM-like Higgs boson mass can get an extra tree-level contribution proportional to the trilinear coupling in superpotential. And the mixing between the SM-like Higgs boson and another lighter Higgs field can help lifting 
the SM-like Higgs boson mass via their mass matrix diagonalization. Because
the large loop contributions from top-stop sector is not needed, the stops can indeed be light as well. 

To evade the current LHC SUSY search constraints, we can consider the following SUSY scenarios
\begin{itemize}
\item{\textbf{R-parity violation}}~\cite{Barbier:2004ez}  If R-parity is broken in SUSY models, the large $\slashed{E}_T$ requirement in most of SUSY searches is no longer fulfilled. So, the corresponding strong bounds can be avoided. 
However, the bounds may remain strong when the R-parity breaking superpotential violates
lepton number but preserves baryon number~\cite{Aad:2014iza, CMS-PAS-SUS-13-010}. 
\item{\textbf{Compressed SUSY}}~\cite{LeCompte:2011cn, LeCompte:2011fh}  When the sparticles and 
Lightest Supersymmetric Particle (LSP) have degenerate masses,  all decay final states of the SUSY events 
will be too soft to be detected at LHC. In this case, the events can only be probed when a hard 
Initial State Radiated (ISR) jet recoils against the sparticle system. Many studies have been carried out for stop~\cite{Ajaib:2011hs,Bi:2011ha}, sbottom~\cite{Alvarez:2012wf}, gluino~\cite{Cheng:2013fma,Bhattacherjee:2013wna}, and Higgsino~\cite{Han:2013usa,Han:2014kaa}. 
\item{\textbf{Displaced SUSY}}~\cite{Graham:2012th} If a neutral sparticle has a decay length around$~\sim $ mm, 
the momentum of its decay products will not point to the primary vertex of the event. 
So, there will be a great possibility that these decay products can not be properly reconstructed. As a result, 
there will be less charged tracks associated to the event, which reduce the trigger efficiency. 
\item{\textbf{Stealth SUSY}}~\cite{Fan:2011yu} In contrast to the compressed SUSY scenario, the missing transverse 
energy ($E^{\text{miss}}_T$) is small because of the softness of the invisible particle itself, not because of 
the cancellation of two invisible particle energies. And then even the ISR jet is not able to produce 
large $E^{\text{miss}}_T$.
\item{\textbf{``Double-invisible" SUSY}}~\cite{Guo:2013asa, Alves:2013wra} If sparticles have multi-body decays into 
more than one invisible particles, both the visible and invisible energies  in the final states will be softened. 
\end{itemize}

Before we propose the heavy LSP SUSY, let us explain the fine-tuning in SUSY models 
and the LHC SUSY search constraints first.
In the Minimal Supersymmetric Standard Model (MSSM), 
the Higgs boson mass is bounded by $Z$ boson mass at tree level, and the top-stop corrections can
lift it to the observed value $m_h \sim 125$ GeV as follows
\begin{align}
 m^2_h \simeq m^2_Z \cos(2 \beta) +  \frac{3 m^4_t}{4 \pi^2 v^2} \left( \text{log}(\frac{M_S^2}{m^2_t}) + \frac{X_t^2}{M^2_S}(1- \frac{X_t^2}{12 M^2_S}) \right)~,~
\label{mhstop}
\end{align}
where $M_S = \sqrt{m_{\tilde{t}_1} m_{\tilde{t}_2}}$, $X_t = A_t - \mu/ \tan \beta$, $v = 174$ GeV,
 and $m_t$ is the running top mass at $m_t$ scale. 
In the large $\tan\beta$ limit, the radiative corrections are required as large as tree-level contribution
\begin{align}
\delta_t^2 \simeq (125)^2 - (91)^2 \simeq 7.34 \times 10^3~.~
\label{cor1}
\end{align}
If the logarithmic term gives the dominant contribution, the average stop mass $M_S$ is required to be above TeV scale. 
However, in the maximal mixing scenario~\cite{Brummer:2012ns} with $X_t \simeq \sqrt{6} M_S$, the third term can 
give dominant contribution to Higgs boson mass without very heavy stop.  So
natural SUSY strongly favours this region. From Eq.~(\ref{mhstop}), we have 
\begin{align}
\frac{3 m^4_t}{4 \pi^2 v^2} \times \frac{X_t^2}{M^2_S}(1- \frac{X_t^2}{12 M^2_S}) \sim \frac{3 m^4_t}{4 \pi^2 v^2} \times 3 \sim 7 \times 10^3~,~
\end{align}
which just satisfies the condition in Eq.~\ref{cor1}.  Then the top-stop sector is mainly constrained by the absence of 
the color breaking vacuum which usually means $|X_t/M_S| \lesssim \sqrt{6}$ and the requirement of
the electric and color neutral LSP ($m_{\tilde{t}_1} > m_{\tilde{\chi}^0_1}$). 
As we will see shortly, in this scenario, the $A_t$ parameter most likely gives rise to the largest fine-tuning, 
even when $\mu$ parameter is relatively large, i.e., about $700$ GeV. 

Concerning the bounds from direct SUSY searches at the LHC, let us take the gluino as an example. 
Gluino is the most copiously produced sparticle because of its strong coupling and high color multiplicity. 
However, searches for direct gluino pair production accompanied by $\tilde{g} \to q \bar{q} \tilde{\chi}$  
decay~\cite{TheATLAScollaboration:2013fha,Chatrchyan:2014lfa} show that if the LSP mass is greater 
than $\sim 500$ GeV, any gluino heavier than this mass can be safely undetected. In order to 
have the bound $m_{\tilde{g}} \gtrsim 1100$ GeV, only $m_{LSP} \gtrsim 400$ GeV is required because of 
the smaller production rate. These bounds become a little bit strong for 
$\tilde{g} \to t \bar{t} \tilde{\chi}$ and $\tilde{g} \to b \bar{b} \tilde{\chi}$ which are the favoured 
decay chains in natural SUSY. The corresponding limit for the LSP mass becomes a little bit higher 
because these searches use b-tag as signal discriminator and then does not depend on the energy of final states 
so much. The gluino can evade all the LHC searches for $m_{LSP} \gtrsim 700$ GeV. And the LSP mass drops to 
$\sim 600$ GeV for gluino with mass larger than $\sim 1100$ GeV~\cite{TheATLAScollaboration:2013tha,Aad:2013wta,CMS:2013jea,CMS:2013ida,Chatrchyan:2013iqa,CMS:2013cfa,Chatrchyan:2013wxa}. The corresponding bounds 
can be relaxed in the realistic MSSM since there will be suppression from branching fractions. Moreover,  
the existence of intermediate on-shell stop or sbottom tends to loose the bounds as well 
if the stop mass is either close to the gluino mass or the LSP mass.  Similar results hold for the first 
two-generation squarks, stops, sbottoms, sleptons, charginos, and neutralinos, respectively.

Based on the fact that all the decay products of sparticles can be softened dramatically when the LSP mass is lifted, 
we propose the heavy LSP SUSY, which can evade the current LHC SUSY search constraints while maintain 
the naturalness condition. 
This scenario is different from the compressed SUSY since only the LSP mass is of concern and 
the mass splitting can be relatively large. Thus, the heavy LSP SUSY can be realized relatively easy
in the Grand Unified Theories (GUTs) as well.
To understand why all the current LHC SUSY search constraints can be escaped, 
we divide the kinematical variables that are used in experimental analysis into three categories. 
In the first category, 
the variables depend quadratically on the LSP mass. For example, $E^{miss}_T$, $H_T$, $m_{\text{eff}}$ and so on. 
We define an energy scale measure $P$ to estimate the typical scale of those variables. 
In the second category,  the variables only depend weakly on the LSP mass. 
For instance, $m_{T_2}$~\cite{Lester:1999tx,Barr:2003rg} 
and some of those razor variables~\cite{Buckley:2013kua} are in this category.  In the third category,
the variables are independent of the LSP mass. e.g., lepton and jet multiplicity, number of b-tagged jets,
 and energy of ISR jet. Most of the scaleless variables  belong to this category. 
Almost all of the current SUSY searches depend heavily on those variables in the first category, 
which make the search sensitivities drop quadratically with increasing LSP mass. As a result, 
we may conclude that the current  LHC SUSY search constraints can be evaded if the LSP mass is 
around 600 GeV or higher. 

On the other hand, in the realistic MSSM, 
we find that the heavy LSP SUSY does not generate more fine-tuning than Higgs boson 
mass. Based on all the current LHC SUSY searches, we systematically study the viable 
parameter space for heavy LSP SUSY, and present four benchmark points which 
realize our proposal concretely. 
What is more, a $\sim 600$ GeV wino LSP can generate the $(g-2)_\mu$ excess as well~\cite{Roberts:2010cj}.
We find the current  LHC sensitivity on those benchmark points can be improved 
if we use the variables in the second category instead of the first category. 

This paper is oragnized as follows. In Section~\ref{kins}, we will discuss the effects of 
the LSP mass on the energy scale of the final states for all three categories in details.  
In Section~\ref{sec:ft}, based on the Barbieri-Giudice measure,
we study the sources of fine-tuning in heavy LSP SUSY. We survey the realistic heavy LSP scenario in the MSSM 
in Section~\ref{sec:mssm}. In Section~\ref{sec:lhc}, after considering the constraints from 
the LHC direct searches, we systematically study the viable parameter space with heavy LSP SUSY in the MSSM and  
propose four benchmark points. The improvements of using the variables in second category instead of 
the first category will be discussed in Section~\ref{sec:imp}.
Finally, some discussions and conclusions are given in Section~\ref{sec:con}.

\section{Kinematic Analysis of HLSP SUSY}
\label{kins}

As one can see in the exclusion plots provided by the ATLAS and CMS 
Collaborations~\cite{TheATLAScollaboration:2013fha, Chatrchyan:2014lfa, TheATLAScollaboration:2013tha, 
Aad:2013wta, CMS:2013jea, CMS:2013ida, Chatrchyan:2013iqa, CMS:2013cfa, Chatrchyan:2013wxa}, 
no bounds can be drawn for the masses of mother sparticles such as gluino, stop, sbottom, squarks, 
electroweakinos, and sleptons when the LSP goes heavy as $500 \sim 600$ GeV, becauses of the 
softness of decay products and small missing energy. Because natural SUSY requires 
light gluino and stops, the heavy LSP SUSY could fit this requirement very well, and 
escapes the LHC SUSY search constraints simultaneously.

To satisfy the natural SUSY requirement and accommodate large enough muon anomalous magnetic moment, 
we can define the Heavy LSP (HLSP) SUSY as follows: gluino is
around $\gtrsim1$ TeV, stop is lighter than 800 GeV, the electroweakinos and sleptons 
are as light as possible, and the LSP is as heavy as 500 -- 600 GeV 
or even heavier.   In fact, the HLSP SUSY can be regarded as a hybrid of the natural 
SUSY and compressed SUSY. In this Section, we will explain how the concept of the HLSP
SUSY works in the collider aspect, via 
both analytical and numerical studies.

\subsection{Kinematic Analysis}

Let us start with the simplest case: a sparticle with mass $M$ decaying into one 
massless SM particle and one LSP with mass $m$.
In the rest frame of mother particle, we have the momentum magnitude of 
two  daughter particles as they have the same momentum from momentum conservation
\begin{equation}
P = \frac{M^2 - m^2}{2M} = \frac{M}{2} (1 - \frac{m^2}{M^2}) ~,~ 
\end{equation}
which is related to the LSP mass $m$ quadratically.   When  $r=m/M$ goes up to 1/2,  $P$ will 
drop to $\frac{M}{2} \times \frac{3}{4}$ -- a shift in 25\%. This $P$ is the most 
important measure in our HLSP SUSY which determines the whole energy scale of the decay process, 
as one will see clearly in the following analyses. The sparticles are mainly produced in the threshold region, so the boost 
effects can be safely ignored in a general analysis.  Assuming uniform angular distribution,
we therefore have the following average transverse energy for the visible SM decay 
product 
\begin{equation}
p_T^{vis} = \langle p_T \rangle \sim \langle \sin \theta \rangle P \sim 
\frac{2}{\pi} P ~,~
\end{equation}
where $\theta$ is the angle between momentum and the beam pipe direction.

Then, we can have the collider observables expressed by the above momentum $p_T^{vis}$ 
\begin{eqnarray}
E_T^{\text{miss}} \sim 2 \langle \cos \phi \rangle p_T^{vis} \sim \frac{4}{\pi} 
p_T^{vis} \sim \frac{8}{\pi^2} P~,~ \label{2bodyeh1} \\
H_T \sim 2 p_T^{vis} \sim \frac{4}{\pi} P ~,~
\label{2bodyeh2}
\end{eqnarray}
where $E_T^{\text{miss}}$ is transverse energy of the LSP, $H_T = \sum^{N_{\text{jet}}}_{i=1} p^i_T$ is 
the scalar sum of jet transverse momenta in the final state, and $\phi$ is half of the angle 
between the two genuine missing momenta.
Therefore, all the observable energy scales are related to the typical energy scale $P$ 
we defined previously. 

If we go from two-body to three-body decay, with the SM decay products massless,
then we can work on it in a similar way as in two-body case. By combining the 
two visible massless particles into a  massive one whose mass is the invariant 
mass the the original two, we obtain the visible energy 
\begin{eqnarray}
p_1^{vis}+p_2^{vis} = \frac{M^2- m^2 + m_{12}^2}{2M} ~,~ \label{ht} \\
\sqrt{(p^{\text{miss}})^2 + m^2} = \frac{M^2- m_{12}^2 + m^2}{2M} ~,~ \label{met}
\end{eqnarray} 
where $p^{vis}_1$ and $p^{vis}_2$ are the scalar momenta of the two SM particles, 
$p^{\text{miss}}$ is the LSP moment, and
$m_{12}=\sqrt{2 p^{vis}_1 p^{vis}_2(1- \cos \theta)}$ is the invariant mass of the system of two 
visible particles with $\theta$ the angle between the two momenta. 
When the two daughters go in the same direction, then we have $m_{12}=0$ and come
back to the massless two-body decay case. Also, we  have $p^{\text{miss}}= \frac{\lambda^{1/2}(M^2, m^2_{12}, m^2)}{2M} < P$
with $\lambda(x,y,z)=x^2+y^2+z^2-2xy-2xz-2yz$. Because the total energy, which is determined by $M$, 
is barely changed, we have a shift in the observables, respect to the 2-body decaying 
case. Thus, we have a smaller transverse missing energy and a larger 
$H_T$ for three body decay from Eqs.~(\ref{ht}) and (\ref{met}). 
The same reasoning can go further. If the SM product is a top quark which has a 
non-zero mass and then would decay, we call the topology as a multi-body decay. In 
this case, the missing energy would decrease further but not change too much
respect to 3-body decay, as will be shown in the numerical analysis. 
Actually, there is a common feature from $n$-body to $(n+1)$-body decay: the more split one has, 
the softer each product particle would be. However, this shift is small in most cases.
It should be noticed that only the final state jets with transverse energy above some thresholds 
are considered at the hadron collider, which makes our above discussion as an approximation. 
As we will see later, the $H_T$ of three body decay is indeed smaller than $H_T$ of 
two body decay in the relatively compressed region. This is mainly because in this region, 
there is great possibility that the softer jets of three body decay have energies below 
the threshold ($p_T < p_{T_{\text{min}}}$) or go outside the detector ($|\eta|> \eta_{max}$). 

Now we have seen that the collider observables such as  $E^{\text{miss}}_T$ and 
$H_T$ depend on the energy measure $P$ linearly. As for $P$, for a 
fixed $M$, it depends on the LSP mass quadratically. So when the LSP mass increases, 
the energy observables decrease faster and faster.


\subsection{Three Categories of Kinematical Variables}

As we have shown in the previous subsection, variables like $E^{\text{miss}}_T$ and 
$H_T$ are proportional to $P$ which decreasing quadratically as increasing 
the LSP mass. 
The strong dependence of current SUSY searches on those variables render those 
searches less sensitive when the LSP mass is heavier. And this is the main 
motivation for HLSP SUSY. We define all those variables, which depend quadratically 
on the LSP mass, as the first category. 

There are some variables which depend on the LSP mass much weaker than $P$. 
Some of those Super-razor variables~\cite{Buckley:2013kua} are defined as 
ratio of two massive variables. As a result, the strong dependence on 
the LSP mass for each of the variable are canceled. 
The endpoint of the distribution of stransverse mass, 
$m_{T_2}$~\cite{Lester:1999tx,Barr:2003rg}, gives the mass scale of a pair produced 
particles, each subsequently decay into visible and invisible sector. $m_{T_2}$ 
is usually used to discriminate the new physics signal because the end point for 
SM processes are tend to be much lower than new physics processes. 
 The $m_{T_2}$ dependence on the particles masses and trial LSP masses is studied analytically 
in Ref.~\cite{Barr:2007hy,Cho:2007dh}.
However, in  realistic searches, the final state can be very complicate, 
which prevent us from reconstructing the event without ambiguity. 
Following the hemisphere algorithm in Ref.~\cite{CMS-PAS-SUS-13-019} from CMS Collaboration, 
we find the LSP mass dependence of the reconstructed $m_{T_2}$ is  weaker than variable $P$. 
The numerical result will be given later.  
The variables, which depend on the LSP mass weaker than variable $P$, are referred as
the second category.

In the third category, we consider the variables that do not depend on the LSP mass at all, as long 
as the energy of final states are above a low threshold. For example, the numbers 
of leptons, jets and b-jets, and ratio $R={E^{\text{miss}}_T}/{m_{\text{eff}}}$. 
We can take variables like the energy of ISR jet in 
this category as well, since it is only related to the energy scale of the process and the mother particle mass.  
This is also the reason why we can use the mono-jet signature to constrain the compressed spectrum. 

In order to have an intuitive view on those three categories, we take a representative variable 
in each category and study its properties numerically, 
namely, $m_{\text{eff}}$, $m_{T_2}$~\cite{CMS-PAS-SUS-13-019} and $n_b$ for sbottom pair production which 
subsequently decay into bottom quark and neutralino.
The effective mass is defined as
\begin{align}
m_{\text{eff}} = E^{\text{miss}}_T + \sum_i p_{T}(j_i) + \sum_j p_T(l_j)~.~
\end{align}
Because jets can be radiated from the initial state and final $b$ jets, there are 
usually more than two jets in the final state. And we do not choose the two 
$b$-jets to reconstruct the $m_{T_2}$ because this will include the interface 
effects between the second and third categories. So we use the algorithm that was
described in Refs.~\cite{0954-3899-34-6-S01,CMS-PAS-SUS-13-019} to reconstruct two 
pseudo-jets out of the multijet events.  After finding the dijet with the largest 
invariant mass, we cluster all other jets to one of the jet in the dijet system 
according to the minimal Lund distance, meaning that jet $k$ is clustered to 
hemisphere $i$ rather than $j$ if
\begin{align}
(E_i - p_i \cos \theta_{ik}) \frac{E_i}{E_i + E_k} \leq  (E_j - p_j \cos 
\theta_{jk}) \frac{E_j}{E_j + E_k}~,~
\end{align}
where $E_i$ and $p_i$ are respectively the energy and momentum magnitude of pseudo-jet $i$, and 
$\theta_{ik}$ is polar angle difference between pseudo-jets $i$ and $k$. 
After we reconstruct the two hemisphere jets, the $m_{T_2}$ for each event is 
calculated by
\begin{align}
m_{T_2} =  \min_{p_T^{\chi(1)} + p_T^{\chi(2)} = p_T^{\text{miss}} } \ [ \  
\max(m_T^{(1)},m_T^{(2)}) \ ]~,~
\end{align}
where $m_T(i) = \sqrt{(m^{\text{vis}(i)})^2 + m^2_\chi + 2(E^{\text{vis}(i)}_T 
E_T^{\chi(i)} - \vec{p}_T^{\text{vis}(i)} \cdot \vec{p}_T^{\chi(i)} ) }$ is the 
transverse mass.
And $n_b$ stands for the number of b-jets in the final state, where the tag 
efficiency and detector effects have been take into account. 

\begin{figure}[h!]
\centering
 \includegraphics[width=0.38\textwidth]{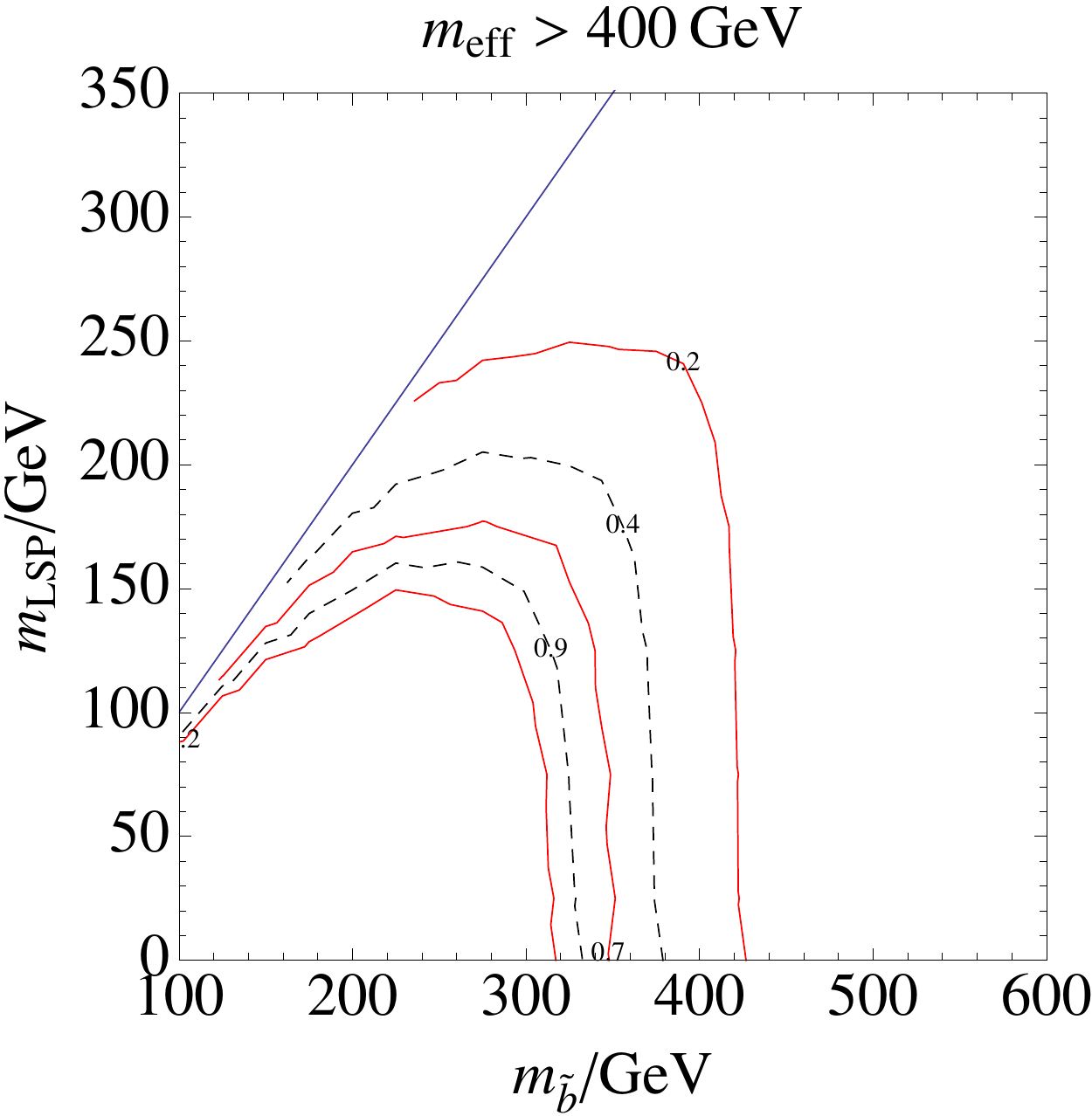}
 \includegraphics[width=0.38\textwidth]{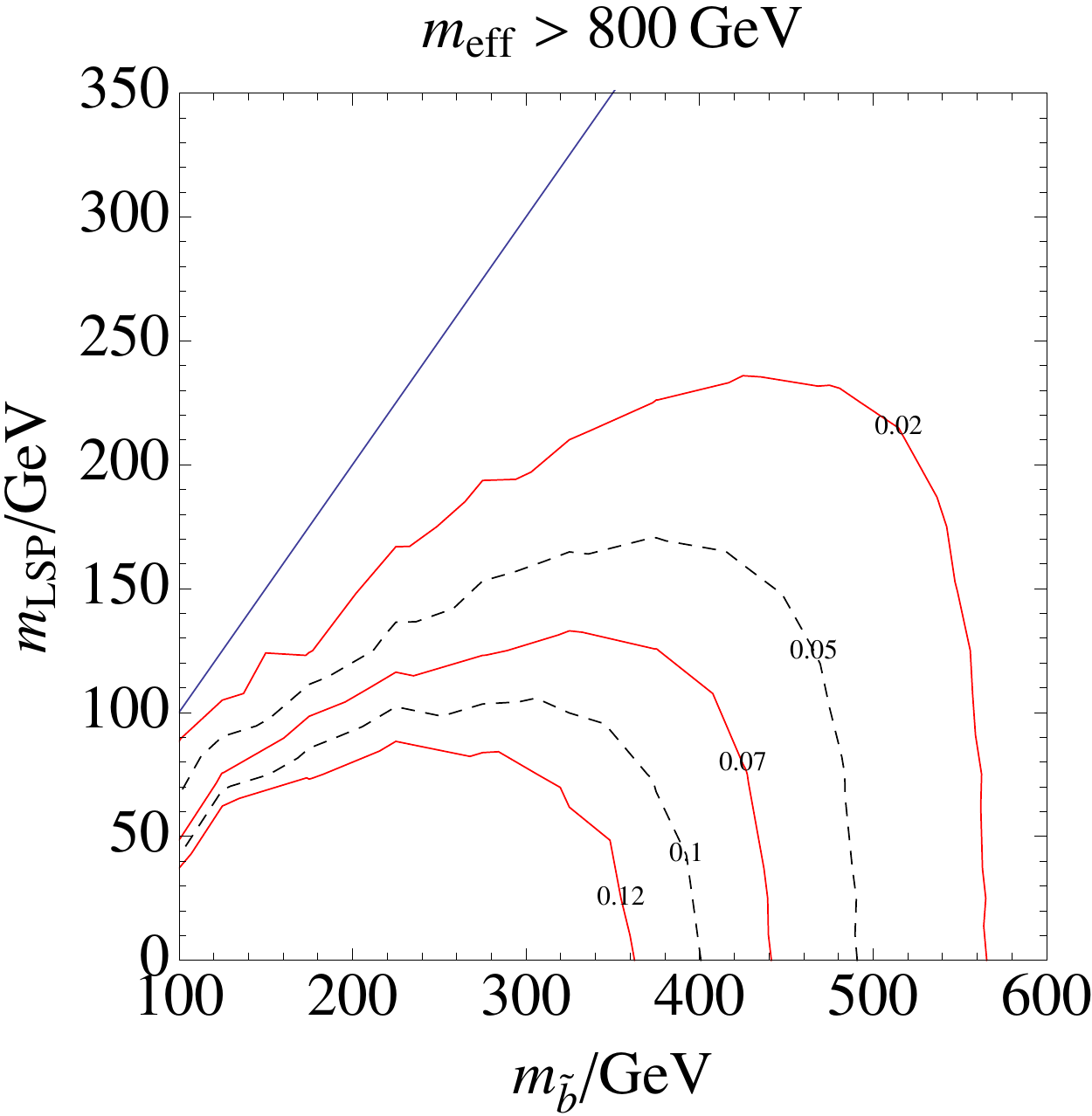} 
 \includegraphics[width=0.38\textwidth]{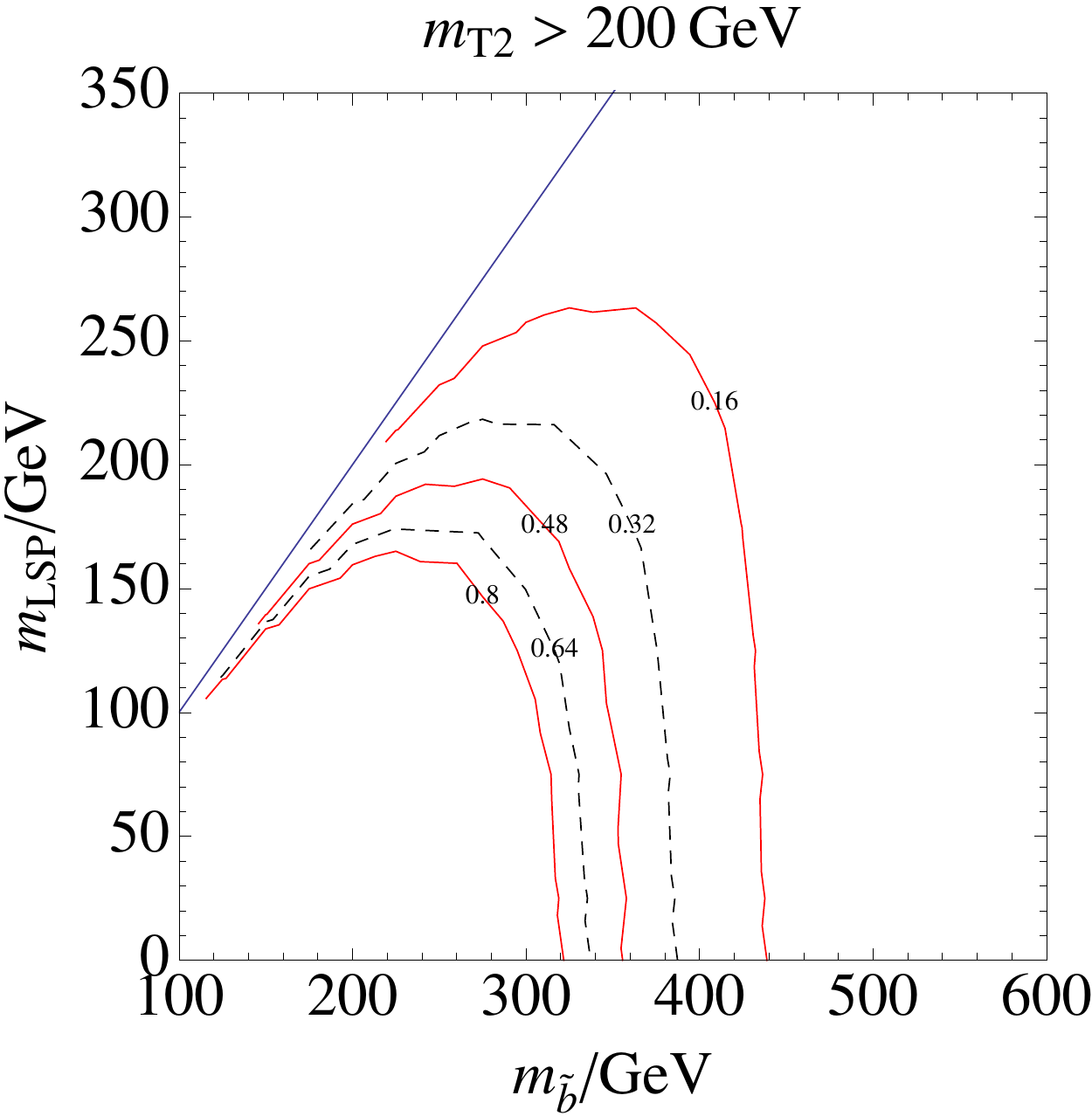}
 \includegraphics[width=0.38\textwidth]{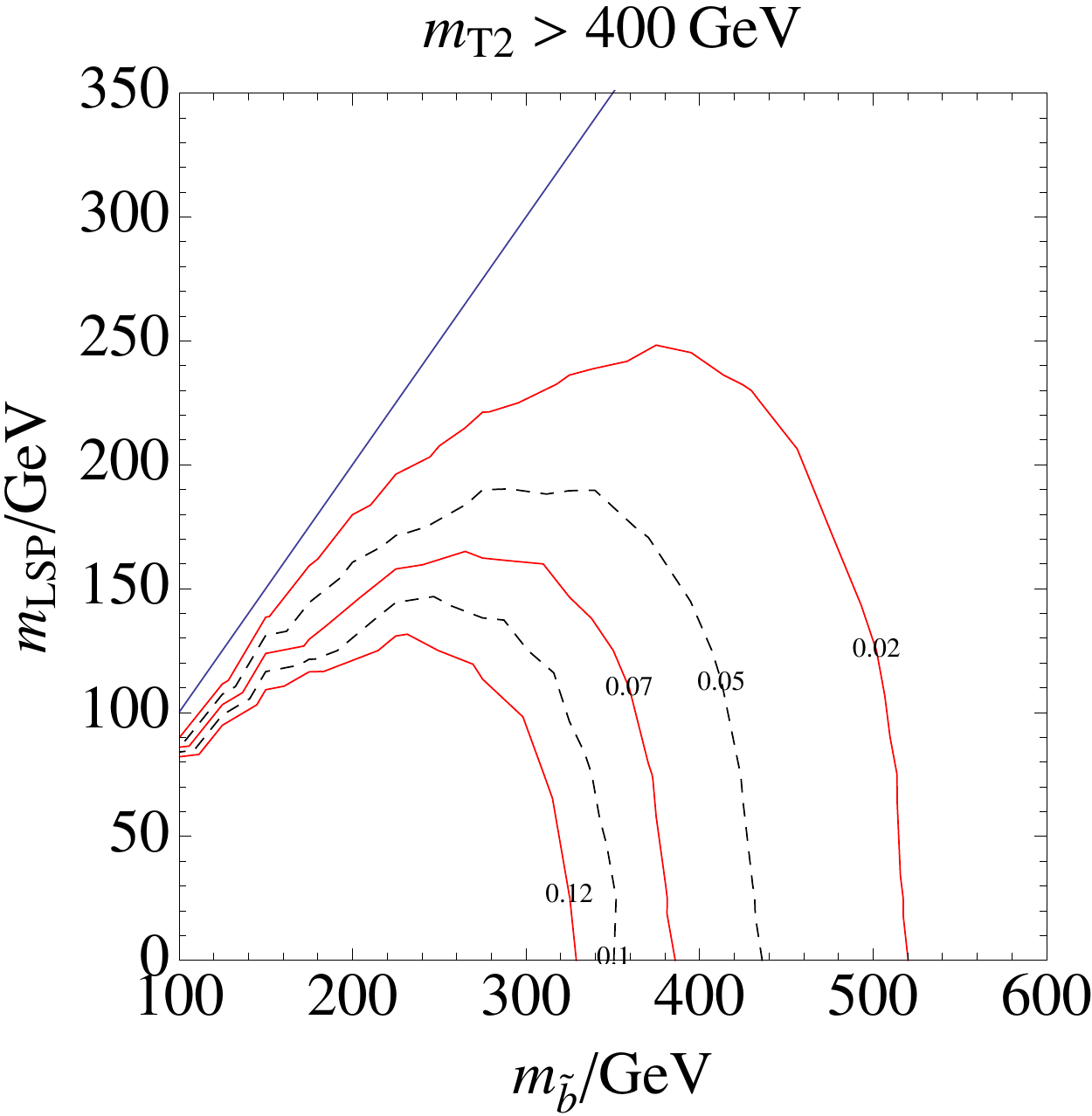} 
  \includegraphics[width=0.38\textwidth]{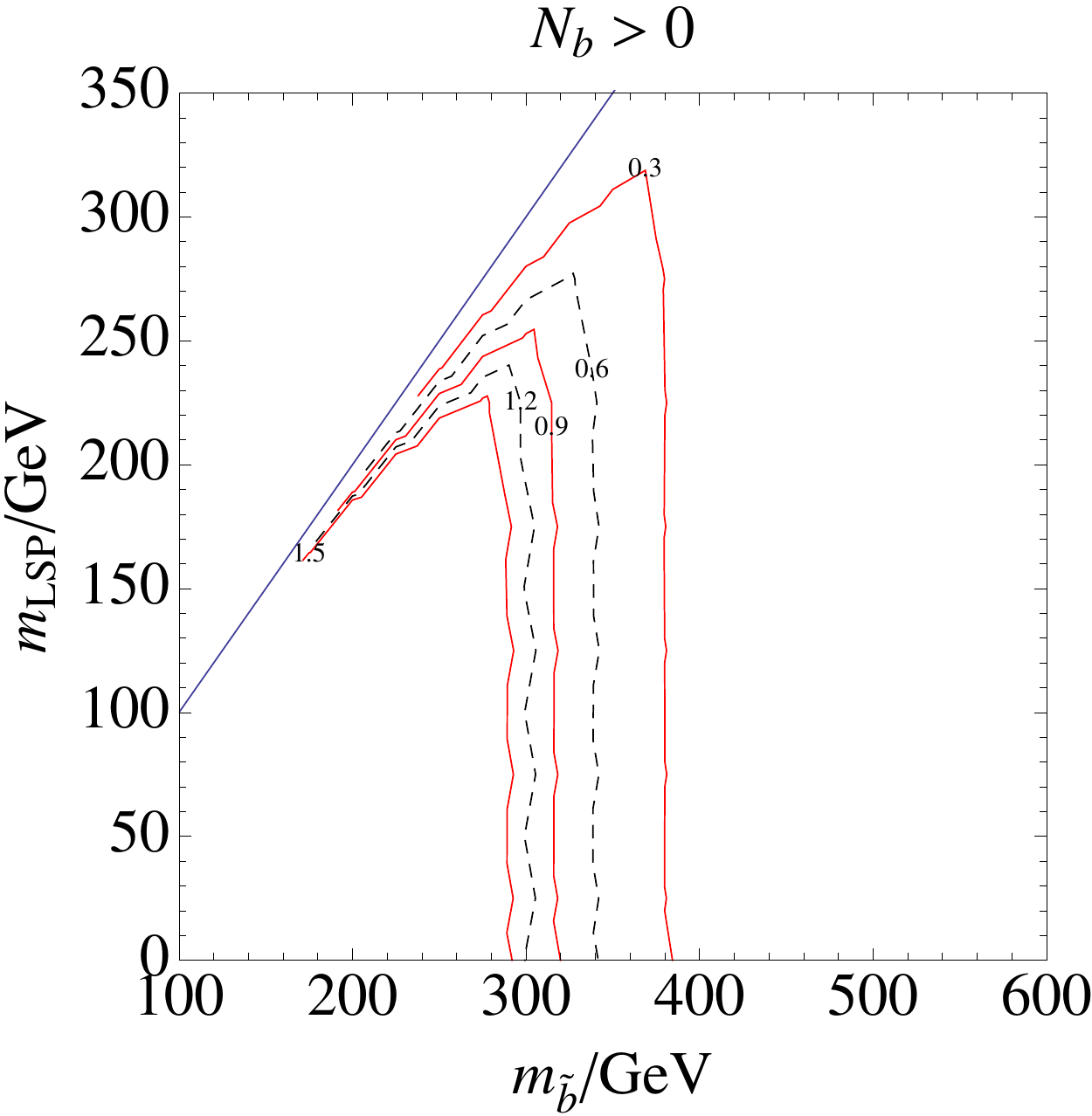}
 \includegraphics[width=0.38\textwidth]{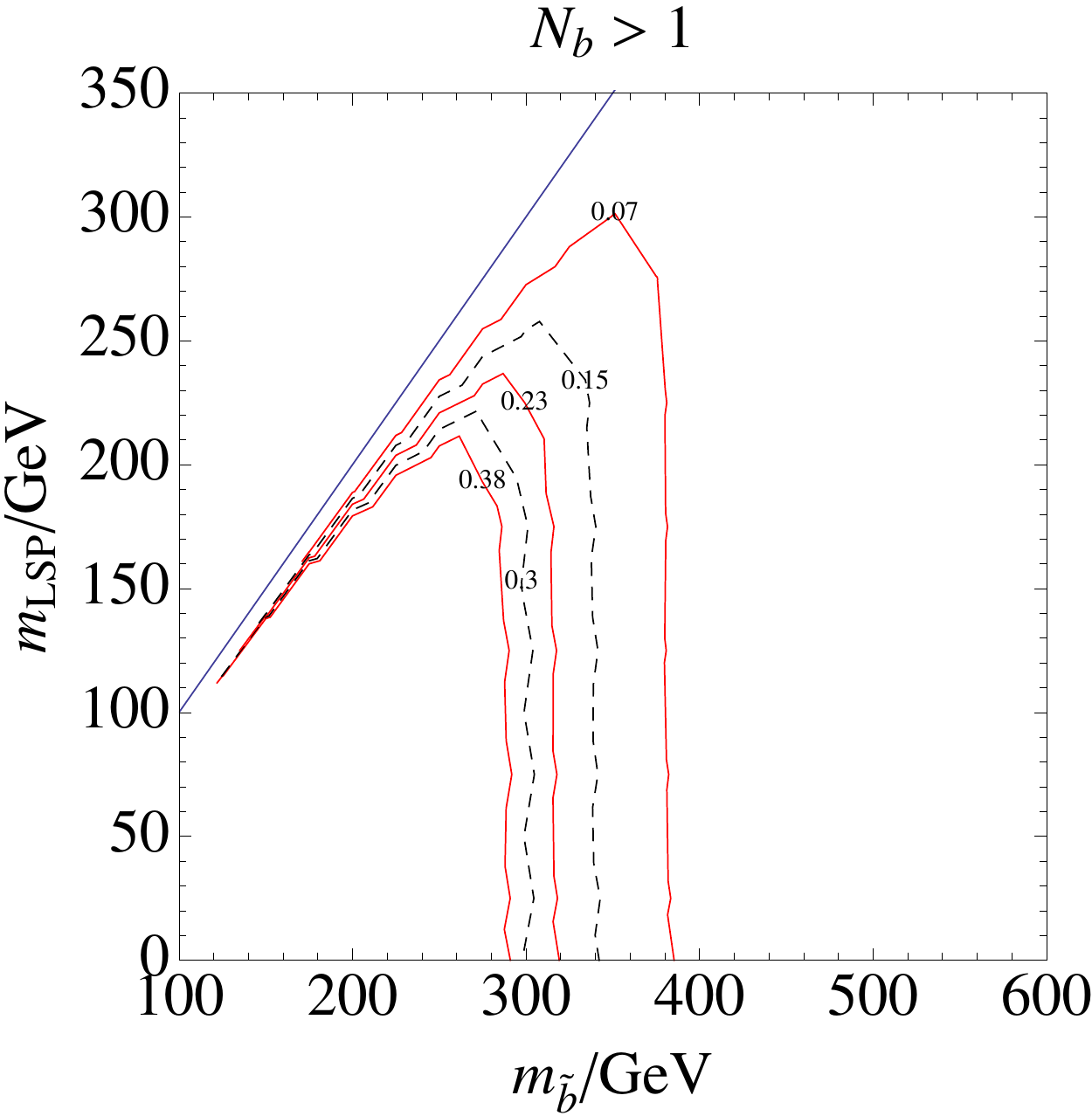} 
 \caption{\label{cate3} $\sigma \times \epsilon$ contour for $H_T$, $m_{T_2}$ 
and $n_b$ cuts(from top to bottom). The blue line shows the $m_{\tilde{b}} = 
m_{\text{LSP}}$. The corresponding cuts are implied as the title for each plot.}
\end{figure}

We use MadGraph5~\cite{Alwall:2011uj} to generate the $\tilde{b}_1 \tilde{b}_1$ 
pair production 
with subsequent decays into the bottom quark and LSP. 
Pythia6~\cite{Sjostrand:2006za} is used for parton showering and hadronization. 
And we use PGS4~\cite{pgs4} for detector simulation, where the default ATLAS 
card has been used. We show the contour of $\sigma \times \epsilon$ for three 
categories in Fig.~\ref{cate3}.

As we can see from the figure, variables of all three categories are not
sensitive to the heavy sbottom and compressed region. This is mainly due to 
small production rate and softness of final states in this region. And all 
contours show steep drop at some high mass region of sbottom because the 
production rate drops exponentially with increasing sbottom mass. 
However, the shape of contours behave very differently for different 
categories. 
The search sensitivity for heavy LSP region becomes stronger from the third category 
to the first category. And putting a more stringent cut will makes the difference 
more significant. The contour is much flatter for the first category, due to
its much strong dependence on the LSP mass. 

\begin{figure}[htb]
\centering
 \includegraphics[width=0.38\textwidth]{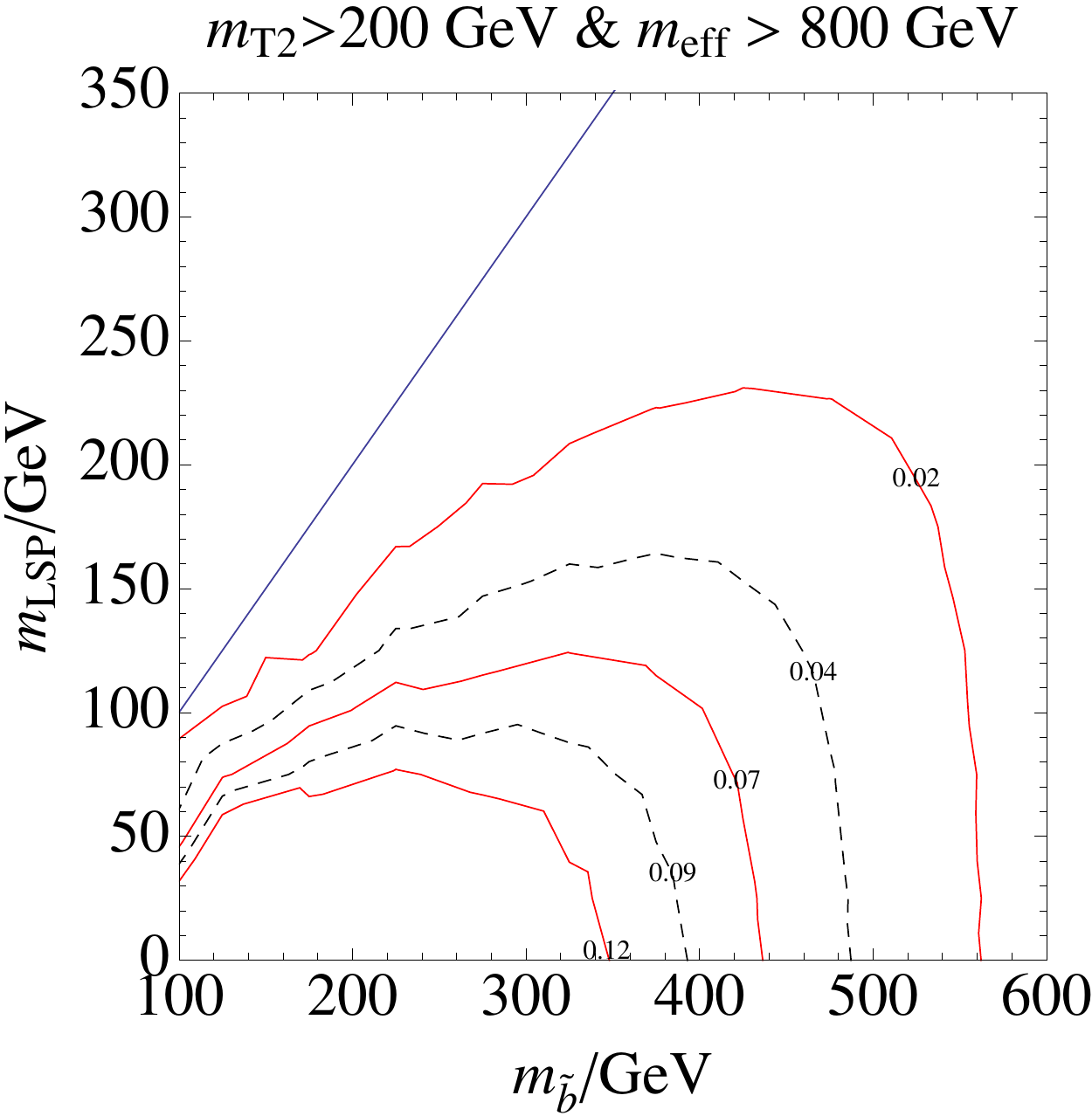}
 \includegraphics[width=0.38\textwidth]{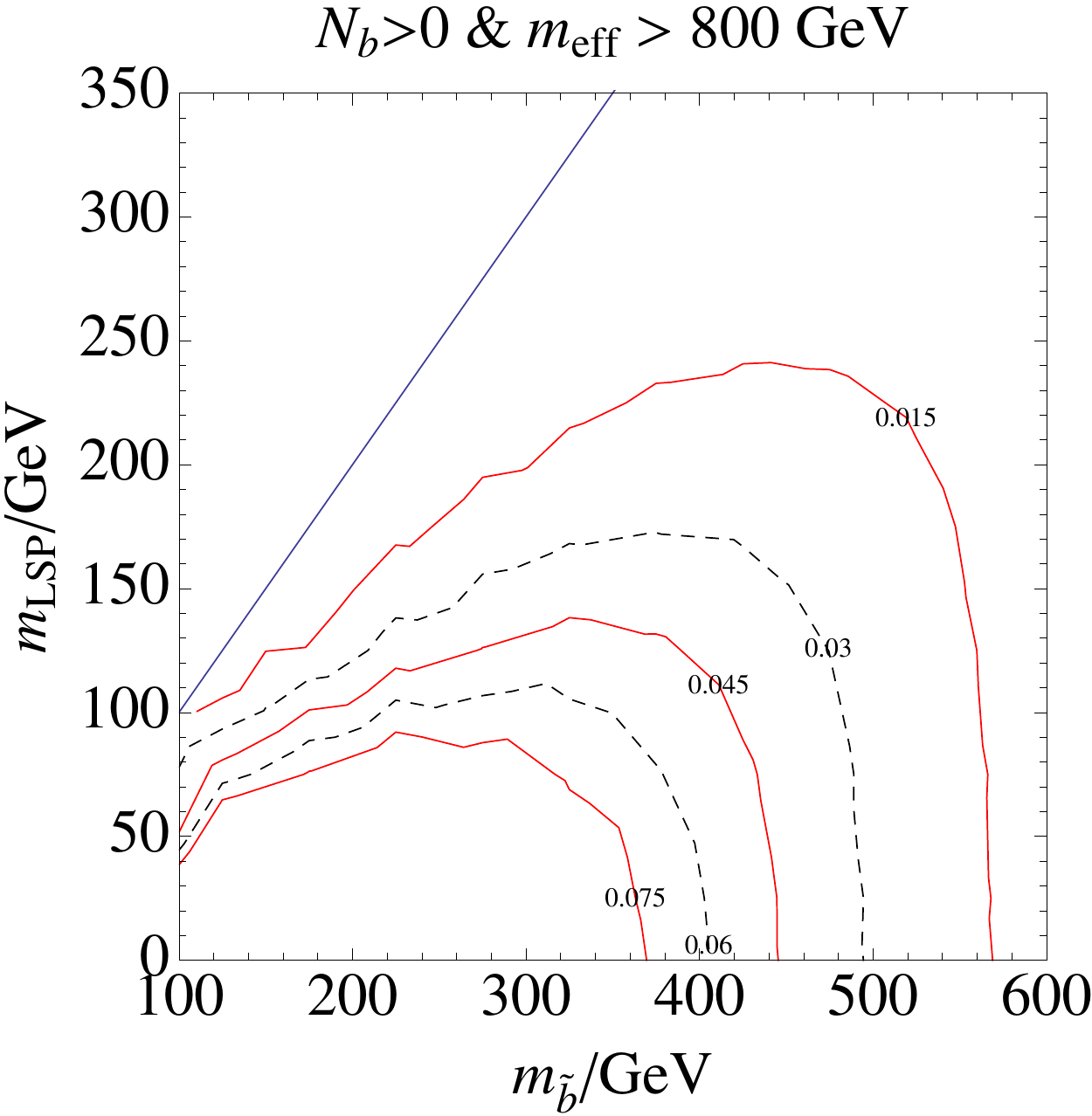} 
 \caption{\label{combc2} $\sigma \times \epsilon$ contour for the 
combined $m_{T_2}$, $m_{\text{eff}}$ (left), $N_b$, and $m_{\text{eff}}$ (right). 
The blue line shows the $m_{\tilde{b}} = 
m_{\text{LSP}}$. The corresponding cuts are implied as the title for each plot.}
\end{figure}

In realistic analysis, we usually need to use the combinative variables in all 
three categories. As a result, the one, which gives the strongest bound, will 
determinate the shape of the  exclusion curve. 
We show the $\sigma \times \epsilon$ contour for
the combined $m_{T_2}$, $m_{\text{eff}}$, $N_b$, and $m_{\text{eff}}$ in Fig.~\ref{combc2}. 
The cut on each variable is chosen as a typical value in experimental analysis. 
From the figure, we can find a typical exclusion curve in 
experimental result which will behave like the contour for those variables in the 
first category. This means the sensitivity to the heavy LSP region is 
relatively weak. And this is case for CMS $m_{T_2}$ analysis~\cite{CMS-PAS-SUS-13-019}. 

\subsection{More on First Category}

To have a more closer look at the first category, we consider three typical 
decay chains for 
gluino as an illustration for two-body decay, three-body decay and massive final 
state respectively: 
1) $\tilde g \to g \tilde \chi$; 2) $\tilde g \to q \bar q \tilde \chi$; 3) 
$\tilde g \to t \bar t \tilde \chi$.  To calculate the LSP mass dependence, 
we fix $m_{\tilde{q}} = 1000$ GeV.

At first, we present the behaviour of averaged $E^{miss}_T$($\langle E^{\text{miss}}_T \rangle$) 
and $H_T$($\langle H_T \rangle$) for different LSP masses and different decay channels 
in Tables~\ref{Tetmiss} and \ref{Tht}, respectively. The averages are obtained with 50,000 events. 
Especially, both collider energies (8 TeV and 13 TeV) are considered 
for decay chain $\tilde g \to q \bar q \tilde \chi$ to show the changes for higher energy.

\begin{table} [htb]
\centering
\scalebox{0.7}{
\begin{tabular}{|c|c|c|c|c|c|c|c|c|c|}
($m_{\tilde g}, m_{LSP}$)& (1000,0)&(1000,100)&(1000,200)&(1000,300)&(1000,400)&(1000,500) & (1000,600)& 
(1000,700) & (1000,800) \\
 $\tilde g \to g \tilde \chi$ at LHC-8 &	554	&	550	& 	534	&	
508	&	480	&435 & 385 & 335 & 282 \\
$\tilde g \to q \bar q \tilde \chi$ at LHC-8 & 	362	& 362	& 363	&  348	& 330	& 299	& 253	& 228	& 178 \\
$\tilde g \to q \bar q \tilde \chi$ at LHC-13 & 	373	&	378	&	374	&	
365	&	341	&307 & 268	& 244	& 204\\
$\tilde g \to t \bar t \tilde \chi$ at LHC-13&	376	&	378	& 374	& 363	& 344	
& 311	& 268\\
\end{tabular}}
\caption{Expected values of $E_T^{\rm 
miss}$ for different LSP masses and different decay channels. All the energy scales are in GeV. 
We display the results for decay chain $\tilde g \to q \bar q \tilde \chi$ at both the LHC-8 and LHC-13 for comparison.}
\label{Tetmiss}
\end{table}

\begin{table}[htb]
\centering
\scalebox{0.7}{
\begin{tabular}{|c|c|c|c|c|c|c|c|c|c|}
($m_{\tilde g}, m_{LSP}$)& (1000,0)&(1000,100)&(1000,200)&(1000,300)&(1000,400)&(1000,500) & (1000,600)& 
(1000,700) & (1000,800) \\
 $\tilde g \to g \tilde \chi$ at LHC-8 & 	1074	& 	1066	&	1045	&	
996	&	936	&851 & 752 & 645 & 534  \\
$\tilde g \to q \bar q \tilde \chi$ at LHC-8 &		1267	& 1231	& 1158	& 1072	& 961	& 850	
& 716	& 598	& 444 \\
$\tilde g \to q \bar q \tilde \chi$ at LHC-13 &  	1391	&	1356	&	1263	&	
1175	&	1070	&945	&	809	& 681	& 523\\
$\tilde g \to t \bar t \tilde \chi$ at LHC-13 &		1381	&	1348	& 1271	& 1168	
& 1066	& 952	& 813\\
\end{tabular}}
\caption{Expected values of $H_T$ for different LSP masses and different decay channels. 
 All the energy scales are in GeV. We display the results for decay chain 
$\tilde g \to q \bar q \tilde \chi$ at both LHC-8 and LHC-13 for comparison. }
\label{Tht}
\end{table}

From these tables we obtain the following results
 \begin{itemize}
 \item  The $P$ measured analysis works very well for all decay channels. 
Both $\langle E^{\text{miss}}_T \rangle$ and $\langle H_T \rangle$ are proportional to 
the corresponding $P$ of each channel and decrease quadratically with increasing LSP mass. 
 \item The ratio of $\langle E^{\text{miss}}_T \rangle$ and $\langle H_T \rangle$ for two-body decay are 
indeed respectively approximated by Eqs.~(\ref{2bodyeh1}) and (\ref{2bodyeh2}), especially 
at the large mass splitting region. There is a little bit discrepancy between 
the theoretical analysis and numerical results, which can be easily explained by 
the boost effects and other realistic facts not taken into account in the ideal calculation. 
From the numerical results, we have $\langle E_T^{\rm miss} \rangle \sim$ P and $\langle H_T \rangle \sim $ 2P.
 \item As we have noted before, $H_T$ will deviate a little bit from theoretical prediction for three body decay 
due to the detector effects. In Table~\ref{Tht}, the $\langle H_T \rangle$ of three-body decay decreases much faster 
than two-body decay when the LSP mass increases, and becomes smaller than the two-body decay 
around $m=\frac{M}{2}$ for different collision energies and different decay channels. 
 \item There are considerable shifts for energy observables from two-body to 
three-body decays from Table~\ref{Tetmiss}. This point would have an impact on the 
collider searches for gluino, since the missing energy cut would lose some 
efficiencies because of the shift downwards. 
 \item From three-body to multi-body, there is no significant change for 
both $\langle E^{\text{miss}}_T \rangle$ and $\langle H_T \rangle$. As a consequence, 
gluinos decaying into any quark types have the similar basic kinematic appearance.
 \item From 8 TeV to 13/14 TeV LHC, both $\langle E^{\text{miss}}_T \rangle$ and $\langle H_T \rangle$ get 
increase because of the increasing hardness of the final states, as expected for the larger Center Mass (CM) energy. 
However, the increase is only mild. 

\end{itemize}

\begin{figure}[ht!]
\centering
 \includegraphics[width=0.4\textwidth]{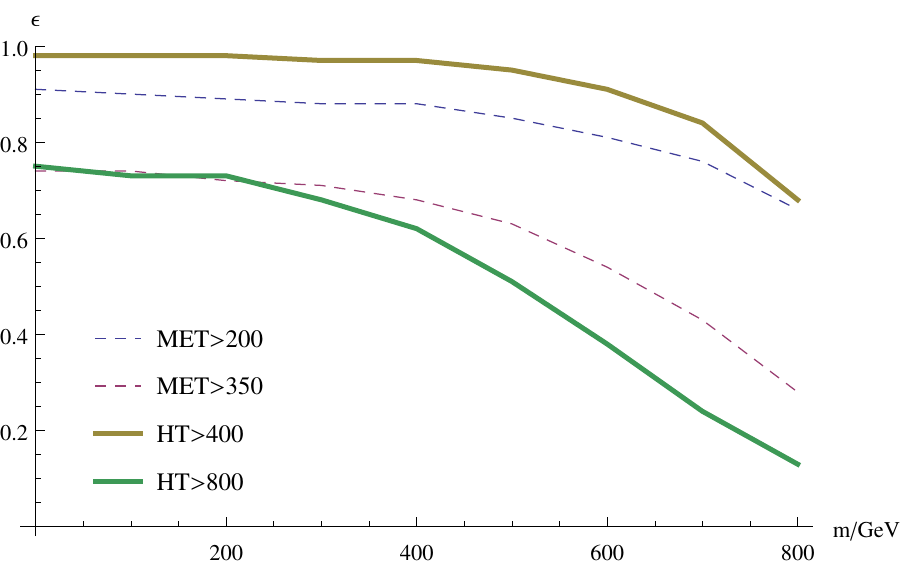}
 \includegraphics[width=0.4\textwidth]{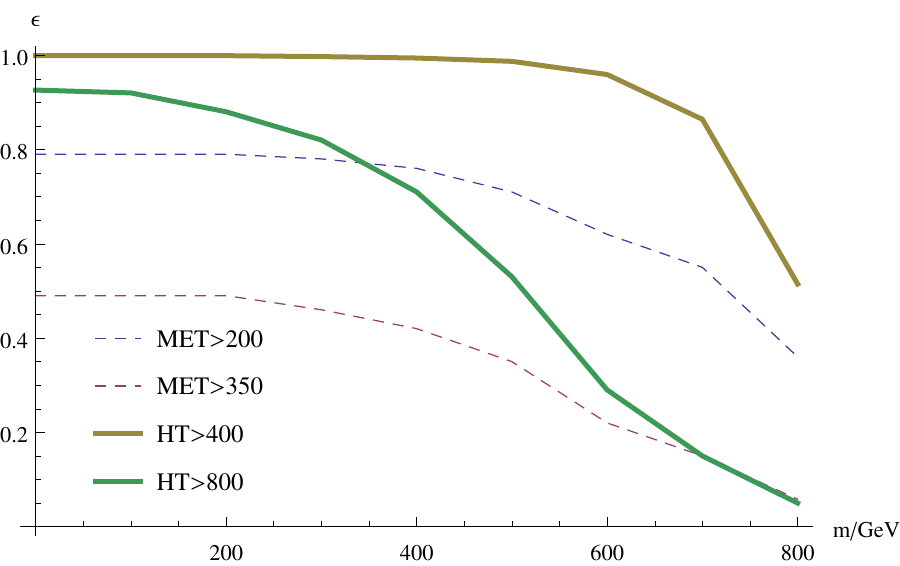}
 \includegraphics[width=0.4\textwidth]{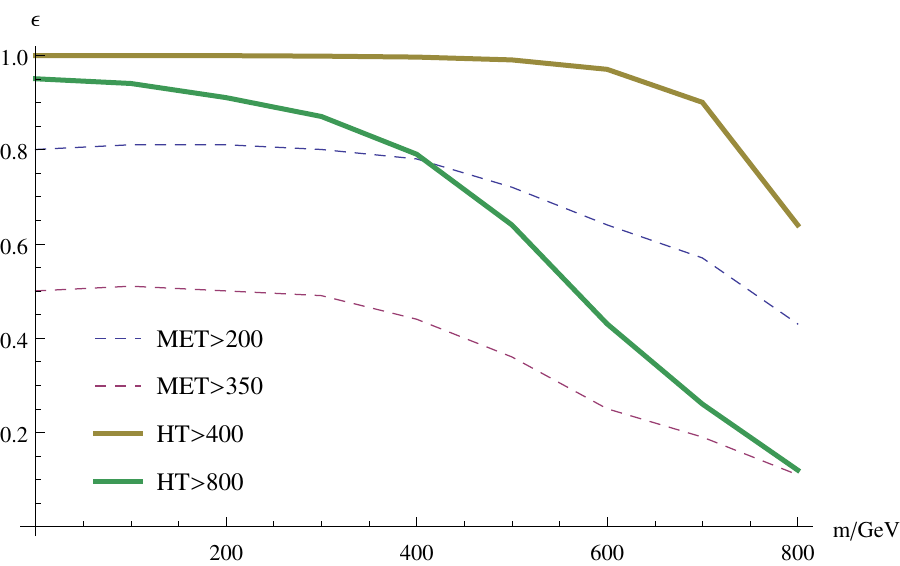}
 \includegraphics[width=0.4\textwidth]{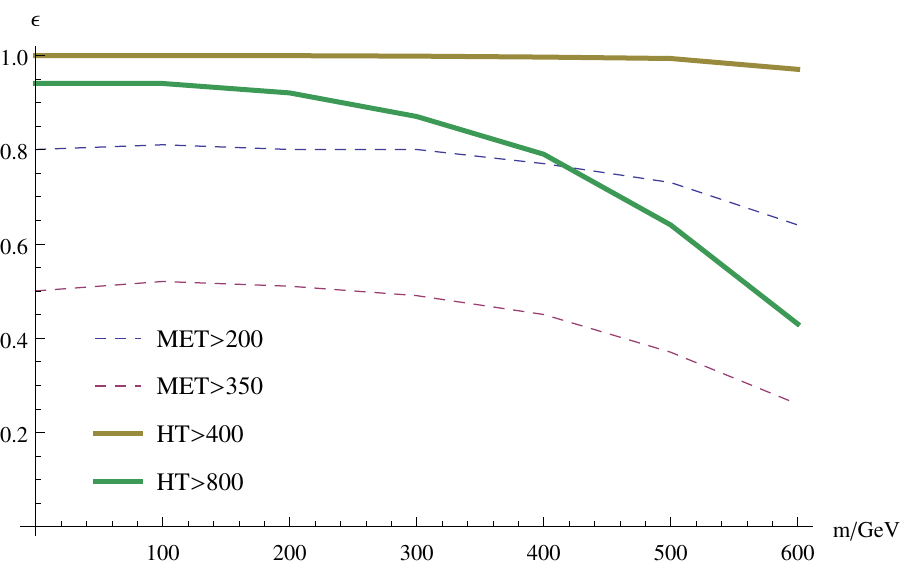}
 \caption{\label{cuteff_m_plot} The cut efficiencies for naive cuts displayed for four cases with $M$ = 1 TeV. 
From left to right and top to bottom, we display two-body decay at 8 TeV LHC, three-body decay at 8 TeV LHC, 
three-body decay at 13 TeV LHC, and multi-body decay at 13 TeV LHC, respectively.}
\end{figure}

Moreover, we want to explore how the cut efficiency is reacting to the LSP mass variance. 
We only investigate very naive cuts for $E_T^{\text{miss}}$ and $H_T$ for a given signal topology. Typical cuts at LHC-8 are employed: $E_T^{\rm miss}>200$ GeV, $E_T^{\rm miss}>350$ GeV, $H_T>400$ GeV, $H_T>800$ GeV. We plot the cut-efficiencies for two-body decay at 8 TeV LHC, 
three body decay at 8 TeV LHC, three body decay at 13 TeV LHC, and multi-body decay at  
 13 TeV LHC with $M=1$ TeV in Fig.~\ref{cuteff_m_plot}. 
 The $P$, hence $E_T^{\rm miss}$ and $H_T$, decrease quadratically with the increase of the LSP mass. 
Thus, the cut efficiency drops much faster when $m$ becomes heavier for all cases. 
From Fig.~\ref{cuteff_m_plot}, we can also conclude that when $r=m/M$  becomes relatively large 
such that $\langle {\mathcal O} \rangle \sim {\mathcal O}_{\rm cut}$(${\mathcal O}$:  observable, ${\mathcal O}_{\rm cut}$: cut for observable ${\mathcal O}$), the cut efficiency for ${\mathcal O}$ experiences a big drop. 
Moreover,  for harder cuts, the LSP mass would play a more important role for cut efficiency. 
As one can see, the multi-body decay case does not differ from the three-body decay case 
(notice the different scaling of $x$-axis in the plots). And the 13 TeV 
results are almost the same as the 8 TeV ones, which makes sense since  we have already read 
that $E_T^{\rm miss}$ and $H_T$ do not change a lot for higher CM energy from 
Tables~\ref{Tetmiss} and \ref{Tht} and the cut energy we employ are relatively low. 
Thus, the cut efficiencies remain the same for higher collider energy. 

As is well-known, the cut efficiency not only depends on the average of ${\mathcal O}$, but also relates to 
the distribution of ${\mathcal O}$. We vary the cut energy for a specific benchmark point to test 
how the cut efficiency changes. It is more like an exploration of the distribution of ${\mathcal O}$. 
The cut efficiency $\epsilon$ is just the p-value here in the simplest case.
Interestingly, having checked several benchmark points, we are led to an empirical formulae shown 
in Table~\ref{law_num} for the 
cut efficiencies, which indicates that the $E^{\text{miss}}_T$ and $H_T$ distributions have a general 
pattern. However, for the realistic collider analysis, a complete Monte Carlo is still needed.
 \begin{table}[ht!]
 \centering
 \begin{tabular}{|c|c|c|c|}
 $E_T^{\text{miss}}$~cut  & $ \langle E_T^{\text{miss}}\rangle$/2 & $\langle 
E_T^{\text{miss}}\rangle$  & 
2$\langle E_T^{\text{miss}}\rangle$ \\
 $\epsilon \sim$ &	0.83	& 0.5	& 0.03-0.05\\
 $H_T$~cut  &  $ 2/3 \langle H_T \rangle$ & $\langle H_T \rangle$ & $5/3 
\langle H_T\rangle$ \\
 $\epsilon \sim$  & 0.82	& 0.45	& 0.05\\
 \end{tabular}
 \caption{The cut efficiencies for different cuts which are scaled by the specific energy scale. We present 
the empirical results here.}
 \label{law_num}
 \end{table}
 
\begin{figure}[ht!]
 \centering
  \includegraphics[width=0.6\textwidth]{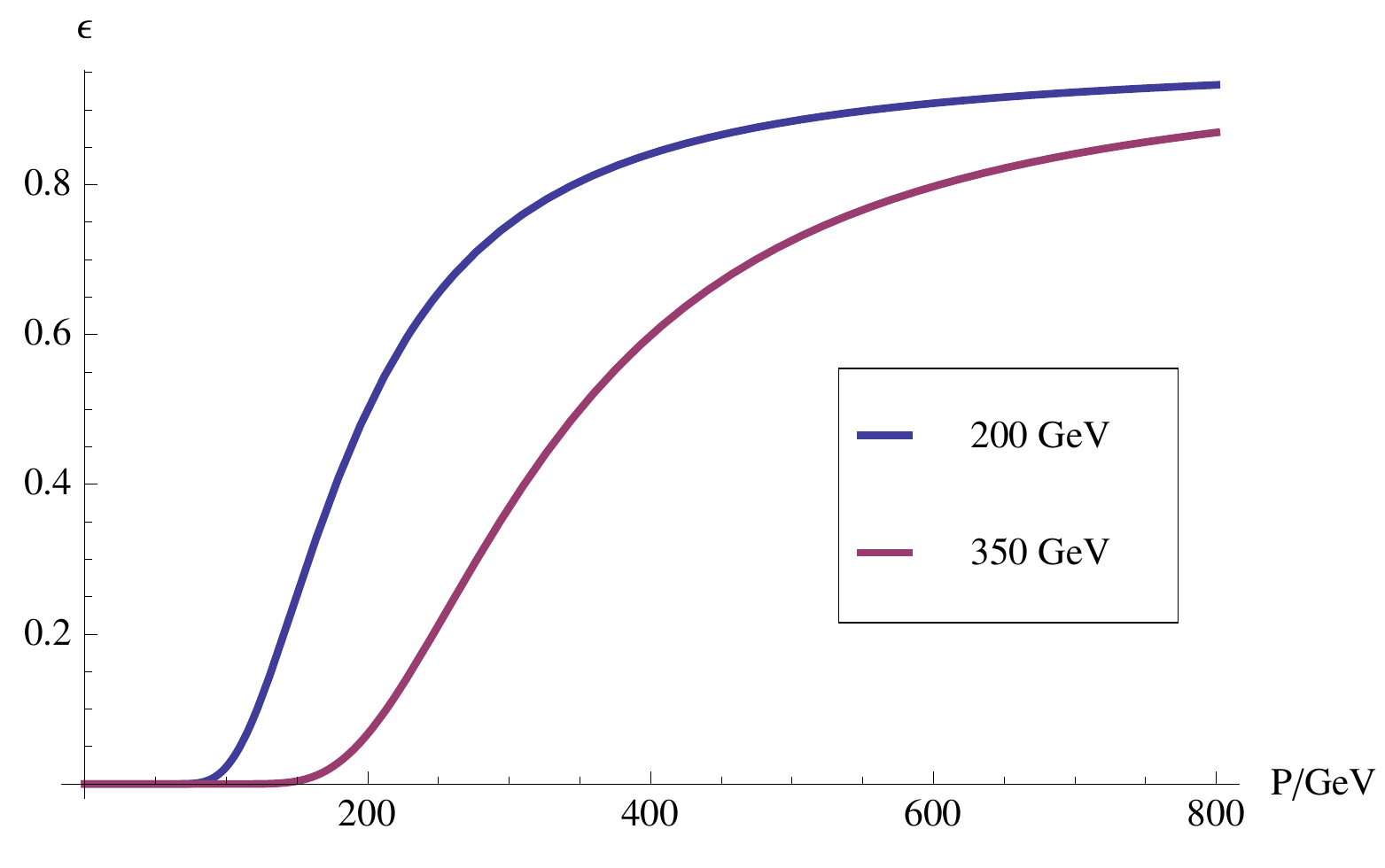}
  \caption{The analytic plot of the cut efficiencies for cut energies 200 and 350 GeV. We are using 
Gaussian distribution for simulation.}
  \label{cuteff_p_plot}
 \end{figure}

 As an approximation, the distributions of $E_T^{\rm miss}$ and $H_T$ can be described by normal distributions. 
For $E_T^{\rm miss}$ the central value is $\langle E_T^{\text{miss}} \rangle$ and $\sigma \sim \langle E_T^{\text{miss}} \rangle/2$, while $H_T$ has a central value of  
$\langle H_T \rangle$ and $\sigma \sim 2 \langle  H_T \rangle/3$. With this approximation along with the estimated $\langle E_T^{\text{miss}} \rangle$ and $\langle H_T \rangle$ previously (Eq.~(\ref{2bodyeh1}) and Eq.~(\ref{2bodyeh2})), we can 
analytically ``guess" the cut efficiencies for any benchmark point. 
Thus, we can encode most of the kinematic information under investigation in 
a single measure P, since knowing P means the distributions of $E_T^{\rm miss}$ and $H_T$ are approximately known.
Then we want to see how the cut efficiency reacts to the change of $P$ for a given cut.
Taking the simplest case: two-body decay under a naive $E_T^{\rm miss}$ cut, 
we show the cut efficiencies for cuts $E_T^{\rm miss}>200$ GeV and $E_T^{\rm miss}>350$ GeV with varying 
$P$ (we have $E_T^{\rm miss} \sim  P$ in this case) in Fig. ~\ref{cuteff_p_plot}, using 
the Gaussian approximation for the distribution of $E^{\text{miss}}_T$.
After a careful comparison one can find that it highly agrees with the realistic one 
in Fig. \ref{cuteff_m_plot} with a simple translation  $ P \to \Delta m = M-m$ 
(see the next subsection for this translation). 
There is a region the cut efficiency drops very fast, which corresponds to our HLSP SUSY
scenario. In this region ($P$ $\in$ [$E_{T{\rm cut}}^{\rm miss}$/2, 2$E_{T{\rm cut}}^{\rm miss}$], or generally $\langle {\mathcal O} \rangle$ $\in$ [${\mathcal O}_{\rm cut}/2, 2 {\mathcal O}_{\rm cut}$]), the cut efficiency changes
violently with the $m$ variance. 

Therefore, in the sense of kinematic analysis, we can redefine the heavy LSP SUSY as follows:
the LSP mass is large enough to give a big impact on the general energy cuts (A visible amount
(e.g. a half) of the cut efficiency drop respects to the massless LSP case.). 
Because the most highly produced sparticle would be gluino, 
we at first consider the HLSP SUSY in the context of gluino production and decay.
Generally for such a heavy LSP, stops and sbottoms are 
always safe because of their low production rate. Thus, at the first sight, the HLSP seems a cut-dependent 
concept. However, with collider energy given, the cut energy scale does not
vary too much due to the limitation of ability of detection.

\section{The Fine-Tuning in the MSSM with a Heavy LSP}
\label{sec:ft}

In previous Sections, we have studied the HLSP SUSY from the kinematic point of view. 
Next, we shall discuss the naturalness of a specific supersymmetric model, the MSSM, 
in which a heavy LSP is imposed.

In the MSSM, the naturalness problem arises from the tadpole equation which determines the $Z$
boson mass $m_Z$
\begin{align}
\frac{m^2_Z}{2} \simeq \frac{m^2_{H_d} - \tan^2 \beta m^2_{H_u}}{\tan^2 \beta -1} - \mu^2 ~,~ 
\label{eq:mz}
\end{align}
where all the parameters are defined at electroweak (EW) scale. 
The following quantity was proposed~\cite{Ellis:1986yg, Barbieri:1987fn} 
to measure the degree of fine-tuning for the above equation
\begin{align}
\Delta_{Z} = \text{max}_i F_i~, ~~~~ F_i = |\frac{\partial \text{ln } m_Z}{\partial \text{ln } p_i}|~,~
\label{eq:ftmeasure}
\end{align}
where $p_i$ are the fundamental parameters at the fundamental scale, for instance, 
the unification scale in the GUTs. 

Because our fundamental theory is usually defined far above the EW scale, we need to express the Higgs soft masses 
$m_{H_u}$, $m_{H_d}$ and $\mu$ in Eq.~(\ref{eq:mz}) in terms of fundamental parameters before 
calculating the degree of fine-tuning in Eq.~(\ref{eq:ftmeasure}).
Previous studies~\cite{Ibanez:1984vq,Lleyda:1993xf} have shown that the soft parameters at low energy scale 
are polynomial functions of the corresponding parameters at fundamental scale, whose coefficients depend on 
the Yukawa couplings and gauge couplings. To derive the polynomial functions by solving the Renormalization 
Group Equations (RGEs), we consider the following MSSM soft terms
\begin{align}
- \mathcal{L}^{\text{MSSM}}_{\text{soft}} & =  -\frac{1}{2} (M_3 \tilde{g} \tilde{g} + M_2 \tilde{W} \tilde{W} + M_1 \tilde{B} \tilde{B} + h.c.) \nonumber \\
 & +[-a_t H_u \tilde{Q} \bar{\tilde{t}} + a_b H_d \tilde{Q} \bar{\tilde{b}} + a_\tau H_d \tilde{L} \bar{\tilde{\tau}} + h.c ] + (\text{first two generation}) \nonumber \\
 & + [\bar{m}^2_Q \tilde{Q}^* \tilde{Q} + \bar{m}^2_L \tilde{L}^* \tilde{L} + \bar{m}^2_t \bar{\tilde{t}}^* \bar{\tilde{t}} + \bar{m}^2_b \bar{\tilde{b}}^* \bar{\tilde{b}} + \bar{m}^2_\tau \bar{\tilde{\tau}}^* \bar{\tilde{\tau}} ] + (\text{first two generation}) \nonumber \\
 & + [m^2_{H_u} H^*_u H_u + m^2_{H_d} H_d^* H_d + (b H_u H_d + h.c.)] ~,~
\end{align}
As a convention, we have defined the following parameters
\begin{align}
a_t \equiv y_t \bar{A}_t , ~~~ a_b \equiv y_b \bar{A}_b, ~~~  a_\tau \equiv y_\tau \bar{A}_\tau ~,~
\end{align} 
where the $y_i$ are the corresponding Yukawa couplings. 

For example, in the MSSM with $\tan \beta =20$~\cite{Abe:2007kf,Feng:2013pwa,Baer:2014ica}, if we choose 
the fundamental scale close to the GUT scale  
$\sim 2 \times 10^{16}$ GeV, without take into account any threshold effect, we can express 
the EW scale $m^2_{H_u}$ and $m^2_{H_d}$ as follows
\begin{align}
m^2_{H_u}  \simeq -2.82 & \bar{M}^2_3 + 0.206 \bar{A}^2_b -0.15 \bar{A}_b \bar{M}_3 - 0.23 \bar{A}_t^2 + 0.48 \bar{A}_t \bar{M}_3 \nonumber \\ 
&  -0.026 \bar{m}^2_{b} -0.42 \bar{m}^2_{Q} -0.33 \bar{m}^2_{t} +0.57 \bar{m}^2_{H_u} + \cdots ~,~ \label{eq:mhu2} \\
m^2_{H_d} \simeq -0.83 & \bar{M}^2_3 - 2.6 \bar{A}^2_b + 1.48 \bar{A}_b \bar{M}_3  + 0.06 \bar{A}_t \bar{M}_3 \nonumber \\ 
&  -0.07 \bar{m}^2_{b} -0.06 \bar{m}^2_{Q} -0.06 \bar{m}^2_{t} +0.85 \bar{m}^2_{H_d} + \cdots ~,~
\end{align}
where the parameters with bar are defined at the GUT scale, the dots denote the ignorable terms, and 
we have used the SM parameters as below
\begin{align}
 &m_{t} =173.4 ~\text{GeV} , ~~~ m_{b} = 4.25 ~\text{GeV}, ~~~ m_{\tau} = 1.777 ~\text{GeV} \nonumber  \\
  & \alpha_1 = 1/58.97, ~~~ \alpha_2 = 1/29.6 , ~~~ \alpha_3 = 1/8.4 ~.~
\end{align}

At the large $\tan\beta$ limit, Eq.~(\ref{eq:mz}) can be rewriten as
\begin{align}
m^2_Z \simeq -2 m^2_{H_u} -2 \mu^2 ~.~
\end{align}
After substituting $m^2_{H_u}$ in Eq.~(\ref{eq:mhu2}), we obtain
\begin{align}
m^2_Z \simeq -2 \mu^2 & + 5.64  \bar{M}^2_3 - 0.412 \bar{A}^2_b +0.3 \bar{A}_b \bar{M}_3 + 0.46 \bar{A}_t^2 - 0.96 \bar{A}_t \bar{M}_3 \nonumber \\ 
&  +0.052 \bar{m}^2_{b} + 0.84 \bar{m}^2_{Q} + 0.66 \bar{m}^2_{t} - 1.14 \bar{m}^2_{H_u} + \cdots ~,~ 
\label{eq:ftgut}
\end{align}
where the small radiative corrections to $\mu$ have been neglected. 
This is the ultimate equation that we will use to estimate the degree of fine-tuning in the MSSM. 
In light of Eq.~(\ref{eq:ftmeasure}), we then calculate the corresponding degree of fine-tuning for each parameter
\begin{align}
 & F_{\mu} \simeq 2 \times \frac{\mu^2}{ m^2_Z},  ~~~ F_{\bar{M}_3} \simeq (2 \cdot 5.64 \bar{M}_3 - 0.96 \bar{A}_t) \times \frac{\bar{M}_3}{2 m^2_Z}  ~,~ \nonumber \\
 & F_{\bar{m}_Q} \simeq 0.84 \times \frac{\bar{m}^2_Q}{{m}^2_Z},  ~~~ F_{\bar{m}_t} \simeq 0.66 \times \frac{\bar{m}^2_t}{m^2_Z} ~,~ \nonumber \\
&  F_{\bar{A}_t} \simeq (2 \cdot 0.46 \bar{A}_t - 0.96 \bar{M}_3) \times \frac{\bar{A}_t}{2 m^2_Z} ~.~
\label{eq:fsoft}
\end{align}


In our heavy LSP SUSY, the LSP is usually required to be above $\sim 600$ GeV so that the current LHC SUSY search
constraints can be evaded.  In order to have all higgsinos heavier than this mass scale, 
$\mu$ should be $\gtrsim 600$ GeV. This will give us the least degree of fine-tuning in the HLSP SUSY, which is 
\begin{align}
F_{\mu} \simeq 87 \sim \mathcal{O}(100) ~.~
\end{align}
In the following discussions, this value will be regarded as a reference degree of fine-tuning 
for natural MSSM with heavy LSP~\cite{Miller:2013jra}. And then
\begin{align} 
\bar{M}_3 \lesssim 380~\text{GeV} 
\label{eq:gluino}
\end{align}
is required if there is no more fine-tuning produced by gluino mass. And then the EW-scale gluino mass 
can be estimated by $m_{\tilde{g}} \lesssim 2.9 \bar{M}_3 \simeq 1.1$ TeV. 
On the other hand, because of the gluino large production rate,
its mass has been excluded up to 1.4 TeV and 1.3 TeV~\cite{TheATLAScollaboration:2013tha,Aad:2013wta,CMS:2013jea,CMS:2013ida,Chatrchyan:2013iqa,CMS:2013cfa,Chatrchyan:2013wxa} when it is decaying into $t \bar{t} \tilde{\chi}$ and $b \bar{b} \tilde{\chi}$, respectively. Even if gluino decays into the first two-generation squarks, 
where the flavour-tag no longer works, the exclusion bounds can still be as high 
as 1.4 TeV~\cite{TheATLAScollaboration:2013fha,Chatrchyan:2014lfa}. 
To have a viable light gluino with mass around 1.1 TeV, in turn we need the heavy LSP to soften the final states 
as discussed in Section~\ref{kins}. This is the reason why the heavy LSP SUSY serves 
as the simplest and most reliable scenario for natural MSSM. 


Similarly, by requiring no more than $\sim \mathcal{O}(100)$ degree of fine-tuning in stop sector, we get 
\begin{align}
 \bar{m}_{Q_3} \lesssim 990~\text{GeV} , ~~~ \bar{m}_t \lesssim 1120~\text{GeV} ,~~~ \bar{A}_t \lesssim 1400~\text{GeV}~.~
 \label{eq:stop}
\end{align}
Note that these values just satisfy the requirements of implementing $\sim 125$ GeV Higgs boson mass in the MSSM. 
As a result, the stop sector remains the main source of fine-tuning  while the LSP mass can naturally be heavy 
in the heavy LSP MSSM. 



\section{Surveying the Heavy LSP MSSM}
\label{sec:mssm}
\subsection{The Parameter Space}

The MSSM is the most well studied SUSY model, since it is not only simple but also can explain 
many new phenomena beyond the SM. The muon anomalous magnetic moment $a_\mu = (g_{\mu}-2)/2$ is one of most 
precisely measured value, which shows more than 3$\sigma$-level discrepancy from the SM.  
The extra contributions from neutralinos/charginos and sleptons/sneutrinos in the MSSM might 
provide an solution to the $(g_\mu-2)/2$ problem. We require the MSSM to have
 the $a_\mu$ within the $3\sigma$ of its theoretical 
prediction~\cite{Jegerlehner:2007xe,Bijnens:2007pz,Czarnecki:2002nt,Hagiwara:2011af}
\begin{align}
4.7 \times 10^{-10} \leq a_\mu \leq 52.7 \times 10^{-10}
\end{align}
when we scan the parameter space. The SM-like Higgs boson in the MSSM should have mass around $125$ GeV as well. 
Considering both the uncertainties from experiments and theoretical calculations, we impose
\begin{align}
 123.5~\text{GeV} \leq m_h \leq 127.5~\text{GeV}~.~
\end{align}
The signal strength of the Higgs boson for all decaying channels are assured to be highly SM-like 
if we are working in decoupling region which is always true in our following discussions. 
The first discovery of branching fraction of $B_s \to \mu^+ \mu^-$ at the LHCb~\cite{Aaij:2012nna}, 
which is very close to the SM prediction, imposes a strong bound on new physics (95\% C.L.)
\begin{align}
2 \times 10^{-9} < \text{Br} (B_s \to \mu^+ \mu^-) < 4.7 \times 10^{-9}~.~
\label{bsmumu}
\end{align}
In the MSSM, the rate of the process is proportional to $\tan^3 \beta$. Because of the relatively large tan$\beta$ we are considering about, Eq.~(\ref{bsmumu}) does give strong bound on our scenario essentially. In contrast, 
we find the branching fraction of $b \to s \gamma$ is always below the measured value after we have imposed 
the constraint from $(g_\mu-2)$. So, we only suppose an additional flavour-violating soft term 
may compensate the value to fit with measurements and will not consider this experiment 
in the following discussions. Also, we will consider the constraints from dark matter relic density 
and dark matter direct detection in the next subsection. 

We work at the GUT scale without implying any unification condition. 
Suspect2~\cite{Djouadi:2002ze} is used to calculate the mass spectra. And all the 
above mentioned experimental values are calculated by micromegas 3.2~\cite{Belanger:2004yn,Belanger:2013oya}. 
The parameters are scanned in the following ranges before we apply any optimizations
\begin{align}
&\tan \beta: [15,40],  ~~ \mu: [500,1000]~\text{GeV},  ~~ M_A: [200,2500]~\text{GeV}, \nonumber \\
&\bar{M}_1: [1200,2500]~\text{GeV},   ~~ \bar{M}_2: [600,1200]~\text{GeV} , ~~ \bar{M} _3: [330,600]~\text{GeV}, \nonumber \\ 
 &\bar{A}_t: [-2300,2300] ~\text{GeV},  ~~\bar{m}_{L_{2,3}}: [400,1000]~\text{GeV},~~ \bar{m}_{e_{2,3}}:[400,1000]~\text{GeV} ,\nonumber \\ 
 & \bar{m}_{Q_3}: [200,1400]~\text{GeV}, ~~ \bar{m}_{U_3}: [200,1700]~\text{GeV}, ~~\bar{m}_{D_3}:[100,1900]~\text{GeV}, \nonumber \\
 & \bar{A}_b:[-2000,2000]~\text{GeV}, ~~ \bar{A}_l = 0 ~\text{GeV}, ~~ \bar{m}_{Q_{2},U_2,D_2}: [1500,3000]~\text{GeV} ~.~
\end{align}

In order to find out the effects of naturalness on the HLSP SUSY,  we have chosen the region which slightly 
wider than the natural SUSY region  that we have discussed in Section~\ref{sec:ft}. 
The gaugino mass ranges at the GUT scale are chosen such that at the EW scale bino and wino 
are within the mass range of $\sim$ $[500,~1000]$ GeV while the mass of gluino is around $[900,~2000]$ GeV. 
Moreover, $\mu$ should be larger than 500 GeV to get a heavy LSP and smaller than 1 TeV 
as required by naturalness. 

\begin{figure}[htb]
\centering
\begin{subfigure}[htb]{0.47\textwidth}
\includegraphics[width=\textwidth]{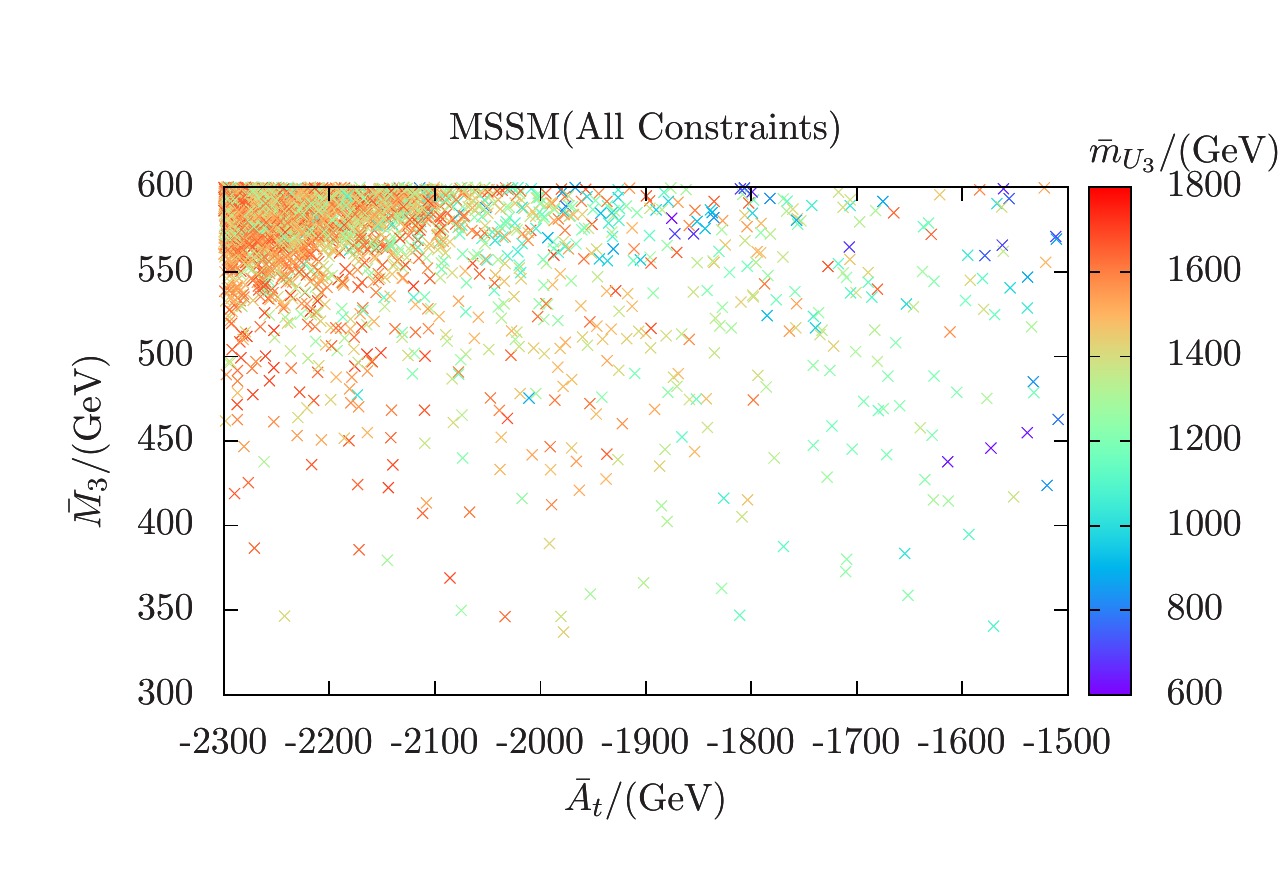}
\end{subfigure}
\begin{subfigure}[htb]{0.47\textwidth}
\includegraphics[width=\textwidth]{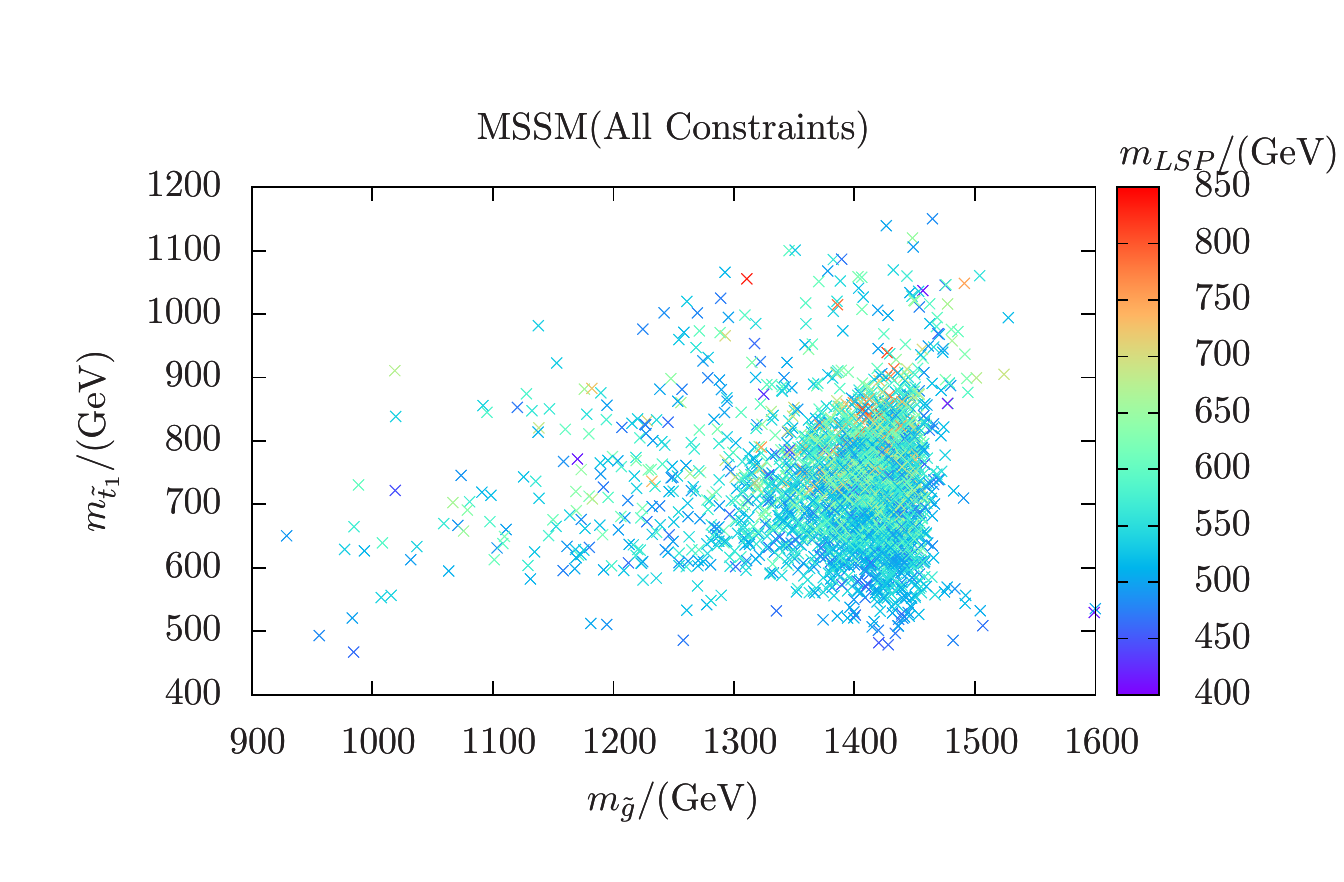}
\end{subfigure}
\caption{The viable parameter space in the heavy LSP MSSM which satisfies all the current constraints 
mentioned in the text except the dark matter constraints.  Left: $\bar{M}_3$ versus $\bar{A}_t$
with varying $\bar{m}_{U_3}$, which give the largest contribution to the fine-tuning measure.  
Right: the gluino mass versus the light stop mass with varying LSP mass.}
\label{atm3}
\end{figure}

To find out the tension between Higgs boson mass and naturalness in heavy LSP SUSY, 
we show the viable parameter space in the left panel of Fig.~\ref{atm3}. 
Due to the heavy Higgs boson mass we have imposed, the upper-left region is highly favoured.  
However, the most natural region is at the lower-right of the 
$\bar{A}_t - \bar{M}_3$ plane, even though only a small parameter space survives in this region. 
Comparing the figure with Eqs.~(\ref{eq:gluino}) and (\ref{eq:stop}), we conclude that 
one of the equations has to be violated in order to get the properly large Higgs boson mass. 
Thus, the heavy LSP SUSY brings no more fine-tuning than the Higgs boson mass. In other words, 
in the HLSP SUSY, the dominant source of fine-tuning is still from Higgs boson mass. 
As for the stop soft masses, a relatively large mass splitting between two stop mass eigenstates 
is preferred to implement the maximal-mixing in stop sector, and we find it is the $\bar{m}_{U_3}$ preferred to be 
the larger one.  So we vary it in the figure as well.

Also, we show the $m_{\tilde{g}}$, $m_{\tilde{t}_1}$ and $m_{\tilde{\chi}^0_1}$ on the right panel of Fig~\ref{atm3}. 
The gluino mass and light stop mass can be as light as $\sim 950$ GeV and $\sim 450$ GeV respectively. 
And the maximal mixing scenario also guarantees that $|\bar{A}_t|$ can remain small ($\sim 1700$ GeV), 
while Higgs boson mass is relatively large. The LSP mass is relatively free as long as it is smaller than 
the light stop mass. 

\subsection{The LSP Properties and Dark Matter Direct Detection}

The heavy LSP can be an interesting dark matter candidate, so  
we will study its properties and direct detection potential. 
Because light wino and higgsino are preferred to enhance the SUSY $(g-2)_\mu$ contribution~\cite{Cho:2011rk},
either wino or higgsino can be the LSP in our scenario.  
To maintain these possibilities, we only require that the LSP be 
one component of dark matter sector. 
So the LSP relic density might be smaller than the Plank measured value 0.1187~\cite{Ade:2013zuv}. 

\begin{figure}[htb]
\centering
\begin{subfigure}[htb]{0.47\textwidth}
\includegraphics[width=\textwidth]{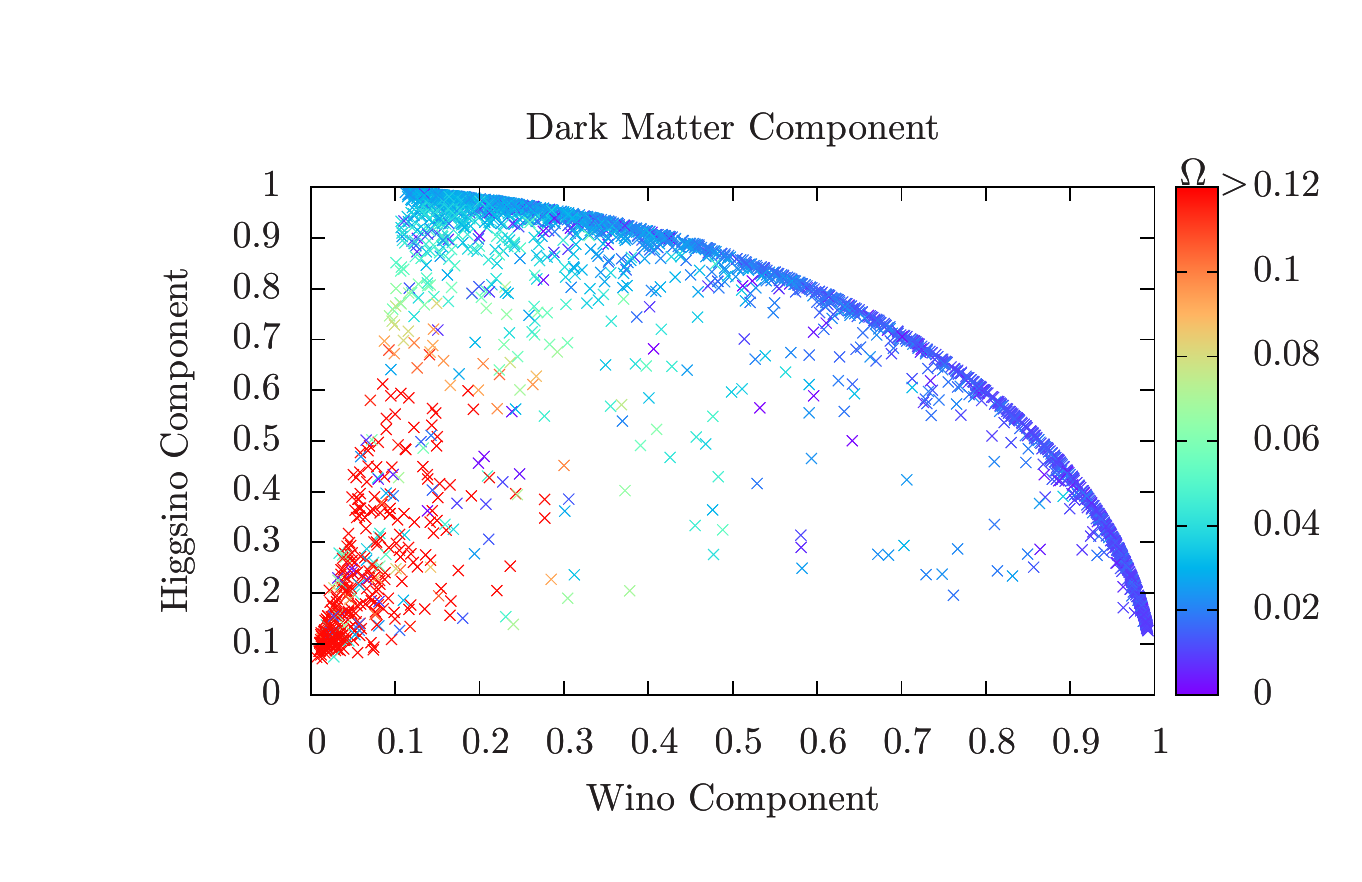}
\end{subfigure}
\begin{subfigure}[htb]{0.47\textwidth}
\includegraphics[width=\textwidth]{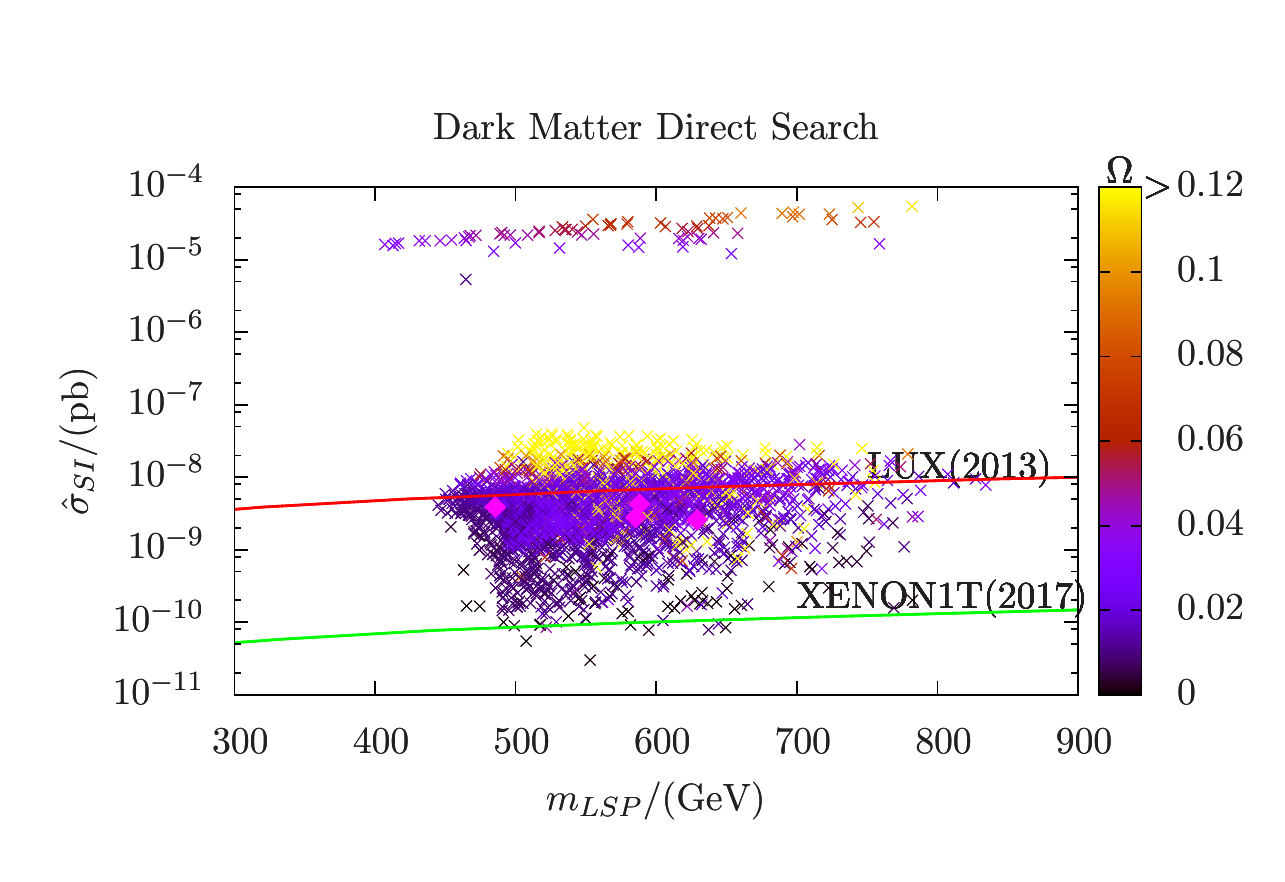}
\end{subfigure}
\caption{Left: The wino and higgsino components of the LSP neutralino.  Right: the LSP mass versus 
the reduced spin independent proton-neutralino collision cross section. The color shows the dark matter 
relic density, where we have set those with $\Omega > 0.1187$ the same color.}
\label{dmp}
\end{figure}

In the left panel of Fig~\ref{dmp}, we show the higgsino and wino components of the LSP neutralino. 
The color shows the relic density, and the red points with $\Omega > 0.1187$ are excluded. 
The major scanned parameter space has
\begin{align}
\text{Wino component}^2 + \text{Higgsino component}^2 \sim 1~.~
\end{align}
Thus, the relic densities for those points are very small, e.g., about one order of magnitude smaller 
than the observed value. However, some points can still have small $\Omega$ while their bino component is large
since the bino LSP can coannihilate with other sparticles, especially stop, to reduce the relic density. 

We show the reduced spin independent dark matter direct detection rate in the right panel of Fig~\ref{dmp}. 
The proton-neutralino collision cross section has been rescaled by
\begin{align}
\hat{\sigma}_{SI} = \frac{\Omega}{0.1187} \sigma_{SI}~,~
\end{align}
which we call it the reduced collision rate. 
The red and green lines show the bounds from the LUX experiment~\cite{Akerib:2013tjd} and 
the expected XENON1T experiment~\cite{Aprile:2012zx}, respectively. And the diamonds are those benchmark points 
which we will discuss later. The yellow points are excluded by  over abundant dark matter. 
As we can see, a few of our viable points have already been excluded by the LUX experiment
and almost all the parameter space are within the reach of the XENON1T experiment. 
Most of the points with large wino and higgsino components satisfy the LUX experimental constraint
because of the small relic density they have. And the others that have relatively large relic density can also 
keep undetected by the LUX experiment due to their large bino component. 

The figure also shows a few points which have very large proton-LSP collision rate. For those points, 
the LSP is $\tau$-sneutrino which has very large spin independent interaction with nucleon though $Z$ 
boson exchange. Their relic density are mainly scaled by their masses, 
ranging from $\sim 0.01$ to $\sim 0.1$ with mass range $\sim$ $[400,~800]$ GeV.

\section{Confronting the LHC Data and Naturalness}
\label{sec:lhc}

The Higgs boson mass is not the only source of fine-tuning problem. 
Even after the realisation of a correct Higgs boson mass in the MSSM, 
either stop or gluino can be very light, as we have shown in Fig~\ref{atm3}. 
However, there are also many direct SUSY searches at the LHC 
which push SUSY above the TeV scale. From discussions in Section~\ref{sec:ft}, 
they might induce a much more serious fine-tuning problem. Unlike the constraint from Higgs boson mass, 
the direct SUSY searches increase the mass bounds on all sparticles simultaneously.  
In this Section, we will study the constraints from the LHC direct SUSY searches and
 find out how heavy the LSP can save the natural MSSM. 

A light stop always exists in natural SUSY models. The searches for light stop are
 one of the main subjects on SUSY phenomenology. The ATLAS and CMS experimentalists  
have carried out many kinds of searches for direct stop production persistently. 
Stop can decay into $b \tilde{\chi}^\pm$ and $\tilde{\chi}^\pm \to W^{(*)} \tilde{\chi}$. 
The searches for final state with b-jets and leptons~\cite{Aad:2013ija,Aad:2014qaa,Chatrchyan:2013xna} 
have excluded the stop in this channel with mass up to about 600 GeV. However, these searches are only 
sensitive when the LSP mass is smaller than $\sim 300$ GeV. The exclusion bound on stop mass
can be higher if $\tilde{t} \to t \tilde{\chi}$, which is $\sim 700$ GeV when considering 
two hadronic decaying tops~\cite{ATLAS:2013cma,CMS:2013cfa} and $640$ GeV when considering 
semi-leptonic decaying tops~\cite{ATLAS:2013pla,Chatrchyan:2013xna}. Both searches are 
heavily rely on the energetic top quarks to suppress the huge $t\bar{t}$ background, 
so they will immediately loose the sensitivity when the LSP mass goes up to $\sim 300$ GeV.

As mentioned in the Introduction, for gluino decaying into the first two-generation quarks and LSP, 
searches for final state with energetic jets and large missing energy~\cite{TheATLAScollaboration:2013fha} 
as well as final state with high jet multiplicity~\cite{Aad:2013wta,Chatrchyan:2014lfa,Chatrchyan:2013wxa} 
will put very strong bounds on gluino mass.  When gluino is decaying through the third generation squarks, 
the final states will contain many b-jets and leptons. So, 
the searches for multi-$b$-jets~\cite{TheATLAScollaboration:2013tha,CMS:2013cfa,ATLAS:2012oqj} and 
lepton plus multi-jets~\cite{CMS:2013ida,Chatrchyan:2013iqa} can 
constrain  these channels. What is more, because gluino is a Majorana particle, the same sign di-lepton (SSDL) 
may show up in the final state. The search for SSDL~\cite{CMS:2013jea} can also impose strong constraint 
for gluino, especially when it decays through top squark. 

We  recast all those analyses closely following the method introduced in Ref.~\cite{Cheng:2013fma}. 
Events are generated by MadGraph5, where Pythia6 and PGS have been packed to implement parton shower, 
hadronization and detector simulation as above. In PGS, we take the $b$-tagging efficiency of 70\%, 
with $c$-mistag and light-jet mis-tag rates of 20\% and 10\%, otherwise specified in corresponding analysis. 
The $M_{T_2}$ variables in some of the analyses are calculated 
by Oxbridge Kinetics Library~\cite{Lester:1999tx,Barr:2003rg}. 
By comparing the upper limit on number of new physics events in each signal region and the number of events 
given by our model, we can tell whether our model is excluded or not. And we define 
the $R^i_{\text{vis}}$ to measure the exclusion potential
\begin{align}
R^i_{\text{vis}} = \frac{\text{Number of events given by our model in signal region } i}{\text{Upper limit for 
new physics events in signal region }i} ~.~
\end{align}
Among all the signal regions, we define
\begin{align}
R_{\text{max}} \equiv \maxr_{i} ( R^i_{\text{vis}} ) , ~~{\rm where}~~ i = \text{all signal regions}~.~
\end{align}
Obviously, the model is excluded only if $R_{\text{max}} > 1$. It has to be noted that we are only using 
the leading order sparticle 
production cross sections throughout the work. Thus, the $R_{\text{max}}$ calculated in this study should be rescaled 
by the corresponding K-factor. 

Because we have already scanned the viable parameter space in the heavy LSP MSSM with Higgs boson mass $\sim 125$ GeV, 
most of our models are consistent with the LHC searches. We propose four benchmark points (BP I, BP II,
BP III, and BP IV) in Table~\ref{bmp4} 
to have a closer look at the heavy LSP SUSY. 

\begin{table}[tbh]
 \begin{center}
\small
  \begin{tabular}{rl}
  \begin{tabular}[c]{|c|c|c|c|c|c|}\hline
  Points   &  BP I   &   BP II  &  BP III & BP IV  \\ \hline  \hline
 $\tan \beta$   &  21.6 & 30.4 & 36.6 &  27.2   \\
 $\mu$ &   796.1 & 603.1  & 504.7 &  594.9 \\
 $M_A$ & 1581.2 & 1499.4 & 1673.3 &  1444.6 \\
 $\bar{M}_1$ & 1806.1 & 2133.7 & 1838.9 &  1978.6 \\
 $\bar{M}_2$ & 798.4 & 973.8 & 796.3 & 1115.4 \\
 $\bar{M}_3$ & 445.1 & 545.8 &  550.3 &  413.3 \\
 $\bar{A}_t$ & -1704.4 & -2243.1 & -2071.4 &  -2108.1 \\
 $\bar{m}_{L_{2,3}}$ & 459.6 & 770.2 & 872.0 &  451.8 \\
 $\bar{m}_{e_{2,3}}$ & 638.4 & 943.9 & 631.1 & 892.1 \\
 $\bar{m}_{Q_3}$ & 556.7 & 569.6 & 789.8 &  1012.6 \\
 $\bar{m}_{t}$ & 1195.0 & 1349.5 &  1483.8 &  1530.9 \\
 $\bar{m}_b$ & 1883.0 & 614.2 & 408.4 &  1652.6 \\
 $\bar{A}_b$ & -16.5 & -641.9 & 771.7 &  -676.4 \\
 $\bar{m}_{Q_{2}}$ & 2015.0 & 2232.7  & 2793.8 &  1887.3 \\
 \hline \hline             
 $m_{H_1}$ &     127.0 &  124.1 & 126.7 &  124.1 \\
 $m_{H_2}$  &   1580.7  & 1499.0  & 1673.5 & 1444.4 \\
 $m_{A}$  &     1581.2     & 1499.4 & 1673.3 & 1444.6 \\
 $m_{H^\pm}$ &   1583.0 & 1501.4  & 1675.8 & 1446.9 \\ \hline
 $m_{\tilde g}$ &  1228.9   & 1341.1 & 1422.0 &  1095.1 \\ 
 $m_{\tilde t_1}$&   754.1 & 661.0 & 717.0 &  845.4 \\
 $m_{\tilde t_2}$&    1125.9 & 1046.5 & 1085.4 & 1120.3 \\
 $m_{\tilde b_1}$&   799.2 & 755.8 & 792.8 & 932.7 \\
 $m_{\tilde b_2}$&   2036.7  & 1100.2 & 936.5 & 1696.1 \\  \hline
   \end{tabular}                                        
                                                        
   \begin{tabular}[c]{|c|c|c|c|c|c|}\hline 
 $m_{\tilde \chi^0_1}$ &  629.2 & 588.1  & 485.6 &  585.4  \\ 
 $m_{\tilde \chi^0_2}$ &   733.3 & 608.6 & 510.8 &  601.8 \\  
 $m_{\tilde \chi^0_3}$ &    798.2 & 819.6 & 675.1 &  843.7 \\ 
 $m_{\tilde \chi^0_4}$ &   827.2 & 923.8 & 781.5 &  924.8 \\ 
 $m_{\tilde \chi^{\pm}_1}$ & 630.2 & 592.8 & 490.8 &  591.1 \\                                                 
 $m_{\tilde \chi^{\pm}_2}$ &  817.6 & 820.6 & 676.6 &  924.3 \\ \hline
 $m_{\tilde s_{L}}$ &  2272.3  & 2534.2 & 3025.0 &  2168.1 \\
  $m_{\tilde s_{R}}$ & 2227.2  & 2479.1 &3005.4 &  2062.1 \\
 $m_{\tilde \nu_L}$&   755.8  & 1023.8 & 969.2 &  920.1 \\
 $m_{\tilde \mu_L/\tilde \mu_L}$&  759.8 & 1026.7 & 972.3 &  923.4 \\
 $m_{\tilde \mu_R/\tilde \mu_R}$& 929.2 & 1278.1& 1081.6 &  1147.3 \\
  $m_{\tilde \nu_\tau}$&   720.9 & 954.4 & 831.6 &  865.0 \\
 $m_{\tilde \tau_1}$&  722.1 & 956.2 &804.6 &  867.1 \\
 $m_{\tilde \tau_2}$&  874.0 & 1166.2 & 849.2 &  1058.4 \\
\hline  \hline                                                                                                  
$\text{Br}_{(B_s \to \mu^+ \mu^-)} / 10^9$& 3.35 & 4.31 & 4.53 &  3.65 \\ 
$\Omega h^2 $    &    0.017     &  0.026   & 0.02 &  0.034 \\
$\sigma_p^{SI}~(pb) / 10^{-9}$& 18.0 & 19.4 & 22.9 &  9.7 \\ 
$(g-2)_\mu / 10^{-10}$& 5.27 & 5.67 & 8.65 &  5.17  \\ \hline
$\text{Br}_{(\tilde g \to \tilde t_i t)}(\%)$ & 44 & 51.4 & 44.2 &  34.2 \\ 
$\text{Br}_{(\tilde g \to \tilde b_i b)}(\%)$ & 56 & 48.6 & 55.8 &  65.8 \\
$\text{Br}_{(\tilde t \to \tilde  \chi^0_i t )}(\%)$ & 0 &0  & 38.4 & 51.4 \\
$\text{Br}_{(\tilde t \to \tilde  \chi^\pm_i b )}(\%)$ & 100 & 100 & 61.6 & 48.6 \\ \hline \hline
$R_{\text{max}}$ & 0.32 & 0.13 & 0.20 &  0.55 \\ \hline
$\Delta_Z$ & 161($\bar{A}_t$) & 279($\bar{A}_t$) & 238($\bar{A}_t$) & 247($\bar{A}_t$)  \\ \hline
  \end{tabular}                                                     
  \end{tabular}
  \caption{\label{bmp4} Four benchmark points for heavy LSP SUSY in the natural MSSM.}
 \end{center}
\end{table}

From Eq~(\ref{eq:fsoft}), the degree of fine-tuning grows dramatically with gluino mass. 
A gluino with mass larger than 1.9 TeV will produce more than the degree of fine-tuning 300. 
All four benchmark points are chosen such that gluino is not giving the largest fine-tuning. 
In other words, gluino is taken as light as possible. And then the maximal mixing in stop sector 
renders $\bar{A}_t$ always gives the largest fine-tuning. Moreover, a relatively light chargino 
is required to give large enough contributions to $(g_\mu-2)$. So, the LSP should be either wino-like 
or higgsino-like, and then the dark matter can have relatively large collision cross section with nucleons. 
The direct search for dark matter may easily exclude those points. However, the dark matter annihilation 
cross section is also considerable. So, the LSP relic abundance  can be much lower than 
the observed value, which makes our LSP only a component of dark matters. As a result, we find 
all four benchmark points are safe for the LUX experiment while within the reach of 
the XENON1T experiment as shown in Fig~\ref{dmp}. 

BP I is the least fine-tuned model with $\Delta_Z = 161$ originating from $\bar{A}_t$. The gluino mass 
gives about 135 for degree of fine-tuning. And the relatively large $\mu$ contributes degree of fine-tuning 
of 153. All those parameters give rise to similar amount of fine-tuning.  
Because gluino is relatively light for this point, a relatively heavy LSP is needed to ensure 
the safety from the LHC SUSY searches. 
We find $R_{\text{max}} = 0.32$ from searching final states with missing energy and at least three b-jets~\cite{TheATLAScollaboration:2013tha}. 
The gluino cascade decays into the LSP either through $\tilde{g} \to t \tilde{t}_1$ or $\tilde{g} \to b \tilde{b}$. 
The stop is 100\% decaying into $\tilde{\chi}^{\pm}_1 b$ where the light chargino is degenerate 
with wino-like LSP and can be effectively regarded as the LSP. The total energy scale can be estimated as 
\begin{align}
E \sim 2 \times \left(\frac{m_{\tilde{g}}}{2}(1-\frac{m_{\tilde{t}/\tilde{b}}^2}{m_{\tilde{g}}^2}) + \frac{m_{\tilde{t}/\tilde{b}}}{2}(1-\frac{m_{\text{LSP}}^2}{m_{\tilde{t}/\tilde{b}}^2}) \right) \sim 1~\text{TeV}~.~
\end{align}
Thus, the effective mass cut ($m_{\text{eff}} \gtrsim 1$ TeV) in Ref.~\cite{TheATLAScollaboration:2013tha} excludes 
a great amount of signal events.  

The stop sector is relatively light in BP II. The large $\bar{A}_t$, which is required by the Higgs boson mass, 
gives the largest fine-tuning $\Delta_Z =279$, and the gluino mass gives the second largest fine-tuning 
$F_{\bar{M}_3} = 202$. However, as discussed above, the heavy LSP ($\sim 600$ GeV) is far beyond 
the current reach ($\sim 300$ GeV). So, there is basically no contribution from the light stop sector. 
Because of the heavier gluino in this point, the search~\cite{TheATLAScollaboration:2013tha} gives 
the $R_{\text{max}} = 0.13$, 
and the energy scale of the point is a little bit higher than BP I
\begin{align}
E \sim 1.2 ~\text{TeV}~.~
\end{align}

Due to the relatively light higgsino LSP in BP III, the muon anomalous magnetic moment can be greatly enhanced. 
As the energy scale of final states at the LHC increases quadratically with decreasing LSP mass, 
the gluino mass should be further lifted to evade the search constraints. 
The sparticle production cross sections at the LHC are dominant by the stop and sbottom pairs. 
And the search for final states with missing energy and two b-jets~\cite{ATLAS:2012oqj} gives $R_{\text{max}}=0.2$.
The energy scale of this point is around
\begin{align}
E \sim 1.5~\text{TeV}~,~
\end{align}
which is significantly higher than BP II. Thus,  its discovery potential is higher than BP II, even with 
heavier sparticles, e.g., gluino, stop and sbottom. 
Again,  $\bar{A}_t$ and $\bar{M}_3$ give the largest  degree of fine-tuning, which are 238 and 206, respectively. 

Finally, BP IV shows the most remarkable feature of the heavy LSP SUSY, whose gluino is very light ($1095$ GeV) 
while the LSP mass is heavy ($585.4$ GeV). Since stop and sbottom are heavy for this benchmark point, 
the LHC searches constrain the gluino mass most.  The corresponding gluino production energy scale is 
\begin{align}
E \sim 900~\text{GeV}~.~
\end{align}
So, the light gluino can still survive the current searches. The most sensitive analysis is still
the search for final states with missing transverse energy and 
at least three b-jets~\cite{TheATLAScollaboration:2013tha}, which gives $R_{\text{max}}=0.55$. 
However,  $\bar{A}_t$ is relatively large for this benchmark point, which render this point suffering from 
the relatively large degree of fine-tuning $\sim$ 247. 

\section{Improving Benchmark Point Searches with $m_{T_2}$}
\label{sec:imp}
As have been shown in Section~\ref{sec:lhc}, the  signal region SR-0l-4j-A in Ref.~\cite{TheATLAScollaboration:2013tha} 
always gives very strong constraint on heavy LSP SUSY, especially  for benchmark points I, II, and IV. 
In this Section, as an example, we will discuss how to improve the search for those benchmark points 
with the variable $m_{T_2}$, which is in the second category. 

According to the cut efficiency on our benchmark points, we classify all those cuts implement in Ref.~\cite{TheATLAScollaboration:2013tha} into three classes
\begin{itemize}
\item{Basic:} $n_l=0$, $p^{j_1}_T >90$ GeV, $\geq 4$ jets with $p_T >30$ GeV, $\geq 3$ b-jets with $p_T >30$ GeV.
\item{High efficiency:}  $\Delta \phi^{4j}_{\min} > 0.5$, $E^{\text{miss}}_T>200$ GeV, $E^{\text{miss}}_T/m^{4j}_{\text{eff}} >0.2$.
\item{Low efficiency:} $m^{4j}_{\text{eff}}>1000$ GeV, $E^{\text{miss}}_T/\sqrt{H^{4j}_T} >16$ GeV$^{\frac{1}{2}}$.
\end{itemize} 

We find cuts of ``high efficiency'' class only remove $\sim 10\%$ of the total events, while those ``low efficiency''
 cuts remove more than half of the total events for our benchmark points. 
Moreover, both of the cuts in ``low efficiency''
 class  belong to  the first category, which render the search relatively insensitive to heavy LSP region, 
i.e., benchmark points. In the following, we will first loosen the cuts in ``low efficiency''
 class to maintain most of the signal events while suppress the background to some extent. 
Then, the $m_{T_2}$ cut is chosen such that the number of backgrounds are suppressed to the same value 
as in Ref.~\cite{TheATLAScollaboration:2013tha}. At last, we compare the number of signal events 
after $m_{T_2}$ cut with those after ``low efficiency'' cuts to find out the degree of improvement. 

The events are generated follow the same procedure as described in previous Section. 
The main backgrounds for our analyses are reducible $t\bar{t}$ and irreducible  $t\bar{t}b$+jets 
where the jets can be either light flavour jets or b-jets and $t\bar{t} + Z/h$ with
 $Z/h$ decays to $b\bar{b}$. Their production cross sections at 8 TeV LHC are given in Table~\ref{bgnum}.  
In order to check the validation of our simulation, the number of background events in control region VR-0l-4j-A 
for both our simulation and experimental simulation are given in Table~\ref{bgnum} as well. 
For our later analyses, we will calibrate the event numbers to those experimental values. 

 \begin{table}[htb]
 \centering
 \begin{tabular}{|c|c|c|c|}  \hline
   &  $\sigma$(8 TeV)/pb &  $N_{\text{VR-1l-4j-A}}$ & $N_{\text{VR-1l-4j-A}}$(EXP) \\ \hline
   $t\bar{t}$ & 222 & 1067 & $840\pm120$ \\ \hline
   $t\bar{t} b$ +jets & 3.36 & \multirow{2}{*}{115} &\multirow{2}{*}{$150\pm120$} \\ 
   $t\bar{t}+Z/h(\to b\bar{b})$ & 0.113 & &  \\ \hline
 \end{tabular}
 \caption{\label{bgnum} The cross sections and validation of background simulation. }
 \end{table}

Our  analysis procedure is described step by step in the following
\begin{itemize}
\item We find more than 80\% of the signal events can be retained if we adjust the ``low efficiency'' 
cuts to $m^{4j}_{\text{eff}}>800$ GeV and $E^{\text{miss}}_T/\sqrt{H_T^{4j}}>10$ GeV$^{1/2}$. 
As a result, the background events are increased by around one order of magnitude. 

\item Follow the hemisphere algorithm in~\cite{CMS-PAS-SUS-13-019}, we construct $m_{T_2}$ 
for each events. And the number of background events can be reduced by imposing 
a cut on $m_{T_2}$ to the same amount as the signal region SR-0l-4j-A in 
Ref.~\cite{TheATLAScollaboration:2013tha}. By counting the generated background events, 
the corresponding $m_{T_2}$ cut is chosen as $m_{T_2}>580$ GeV. 

\item For signal processes, we first apply the ``basic'' and ``high efficiency'' cuts.  
Then, the selected events will go through two different analysis.  Firstly, those 
``low efficiency'' cuts are applied. The corresponding signal event number is denoted 
as $N^0$. The second way is to apply the adjust cuts: $m^{4j}_{\text{eff}}>800$ GeV 
and $E^{\text{miss}}_T/\sqrt{H_T^{4j}}>10$ GeV$^{1/2}$, together with $m_{T_2}>580$ GeV. 
The corresponding number of signal event is $N$. 

\item The number of background events are the same for those two different analyses. 
So, the improvement on signal significance can be approximated 
by $\mu = \frac{N/\sqrt{B}}{N^0/\sqrt{B}}=\frac{N}{N^0}$.

\end{itemize}

We give the $N$ and $N^0$ out of 50,000 generated events for each benchmark points 
in Table.~\ref{nn0bp3}. As we can see, there is indeed an improvement by using harder cut 
on $m_{T_2}$ instead of $m_{\text{eff}}$ and $E^{\text{miss}}_T$. From the table, we can also 
conclude that a greater improvement can be achieved for model with smaller $P$, 
since the ``low efficiency'' cuts become more stringent on those models. 

 \begin{table}[htb]
 \centering
 \begin{tabular}{|c|c|c|c|}  \hline
    & $N^0$ & $N$ & $\mu$ \\ \hline
   BP I  & 1040 & 1342 & 1.3 \\
   BP II & 257 &  312  & 1.2 \\
   BP IV & 1853 & 2842  & 1.5 \\ \hline
 \end{tabular}
 \caption{\label{nn0bp3} The number of signal events for benchmark points I, II and IV 
after two different analyses as described in the text. The total numbers 
of generated events are 50,000 for all benchmark points.}
 \end{table}

\section{Discussions and Conclusion}
\label{sec:con}

We have proposed the heavy LSP SUSY, which can escape the LHC SUSY search 
constraints while preserve the naturalness in the MSSM. Interestingly, all the 
other experimental constraints can be satisfied as well. 

To understand why heavy LSP SUSY can avoid the LHC bounds, according to the 
different dependences on the LSP mass,
we divide the kinematical variables that were used in experimental analyses 
into three categories and studied each of their search sensitivities numerically. 
The sensitivity for the heavy LSP region drops dramatically from the third category to 
the first category. And the variables with weaker sensitivity will determinate the 
shape of exclusion curve in an analysis with combined variables in 
three categories. 
In the first category, 
the $E^{\text{miss}}_T$ and $H_T$ are proportional to $P$ with some coefficients 
depending on the decay modes. 
By assuming their distributions, we can naively estimate the cut efficiencies 
for experimental analyses. 
Thus, we have explicitly shown how the heavy LSP SUSY does work.

In addition, we have considered the naturalness problem for the heavy LSP SUSY 
in the MSSM. 
Using the Barbieri-Giudice fine-tuning measure, we studied the degree of 
fine-tuning produced 
by the GUT-scale parameters in details. 
We found that in the MSSM the 126 GeV Higgs boson mass requirement is always the 
dominant source of 
fine-tuning and then the presence of heavy LSP will not introduce any more 
fine-tuning.  

To realize the heavy LSP SUSY in the natural MSSM, we have scanned the viable 
parameter space.
Based on the collider search sensitivity, we proposed four benchmark points to 
illustrate 
the heavy LSP SUSY. For these benchmark points, the trilinear $\bar{A}_t$ term always gives
the largest fine-tuning, as expected from the stop contributions to 
Higgs boson mass. And gluino can be as light as 1.1 TeV 
with $m_{\text{LSP}} \sim 600$ GeV while keep undetected by all the LHC SUSY 
searches. Meanwhile, 
the existence of light wino or higgsino can explain the $(g_\mu-2)/2$ excess 
within 3$\sigma$ level. 

Almost all of the current SUSY searches mainly rely on either large 
$E^{\text{miss}}_T$ or total energy scale of the events, which are belong to the 
first category in this work. So,  their exclusion bounds are relatively weak in 
the heavy LSP region, even when the mass splitting is relatively large. 
In order to improve the SUSY search on heavy LSP scenario, one should adopt an 
analysis that  mainly depends on the variables in the second category or even in 
the third category. An simple example in this direction was given in this work as well.  
The search sensitivities on those benchmark points can indeed be improved 
by using $m_{T_2}$ instead of $m_{\text{eff}}$.

\section*{Acknowledgements}

This work is supported in part by by the Natural Science
Foundation of China under grant numbers  10821504, 11075194, 11135003, 11275246, and 11475238,
and by the National Basic Research Program of China (973 Program) 
under grant number 2010CB833000.

\bibliography{hlsp}
\bibliographystyle{utphys}

\end{document}